\documentclass[twocolumn,tightenlines,showpacs,prd,floatfix,preprintnumbers,amsmath,amssymb,nofootinbib,superscriptaddress]{revtex4}
\usepackage{graphicx}
\usepackage{subfigure}
\usepackage{color}
\usepackage{psfig}

\newcommand{\be}{\begin{equation}}
\newcommand{\ee}{\end{equation}}
\newcommand{\ba}{\begin{eqnarray}}
\newcommand{\ea}{\end{eqnarray}}
\newcommand{\msun}{M_{\odot}}

\newcommand{\hn}{\hat{n}}
\newcommand{\hm}{\hat{m}}
\newcommand{\ho}{\hat{\Omega}}
\newcommand{\hp}{\hat{p}}

\def\ltsima{$\; \buildrel < \over \sim \;$}
\def\simlt{\lower.5ex\hbox{\ltsima}}
\def\gtsima{$\; \buildrel > \over \sim \;$}
\def\simgt{\lower.5ex\hbox{\gtsima}}

\begin{document}

\title{Measuring the parameters of massive black hole binary systems with Pulsar Timing Array observations of gravitational waves}

\author{Alberto Sesana} 
\email{alberto.sesana@aei.mpg.de}
\affiliation{Albert Einstein Institute, Am Muhlenberg 1 D-14476 Golm, Germany}
\affiliation{Center for Gravitational Wave Physics, the Pennsylvania State University, University Park, PA 16802, USA}
\author{Alberto Vecchio} 
\email{av@star.sr.bham.ac.uk}
\affiliation{School of Physics and Astronomy, The University of Birmingham, Edgbaston, Birmingham, B15 2TT, UK}

\date{\today}

\begin{abstract} 
The observation of massive black hole binaries (MBHBs) with Pulsar Timing Arrays (PTAs) is one of the goals of gravitational wave astronomy in the coming years. Massive ($\simgt 10^8\,\msun$) and low-redshift ($\simlt 1.5$) sources are expected to be individually resolved by up-coming PTAs, and our ability to use them as astrophysical probes will depend on the accuracy with which their parameters can be measured. In this paper we estimate the precision of such measurements using the Fisher-information-matrix formalism. For this initial study we restrict to ``monochromatic" sources, \emph{i.e.} binaries whose frequency evolution is negligible during the expected $\approx 10$ yr observation time, which represent the bulk of the observable population based on current astrophysical predictions. In this approximation, the system is described by seven parameters and we determine their expected statistical errors as a function of the number of pulsars in the array, the array sky coverage, and the signal-to-noise ratio (SNR) of the signal. At \emph{fixed} SNR (regardless of the number of pulsars in the PTA), the gravitational wave astronomy capability of a PTA is achieved with $\approx 20$ pulsars; adding more pulsars (up to 1000) to the array reduces the source error-box in the sky $\Delta\Omega$ by a factor $\approx 5$ and has negligible consequences on the statistical errors on the other parameters, because the correlations amongst parameters are already removed to a large extend. If one folds in the increase of coherent SNR proportional to the square root of the number of pulsars, $\Delta\Omega$ improves as $1/\mathrm{SNR}^2$ and the other parameters as $1/\mathrm{SNR}$. For a fiducial PTA of 100 pulsars uniformly distributed in the sky and a coherent SNR = 10, we find $\Delta\Omega\approx 40\,\mathrm{deg}^2$, a fractional error on the signal amplitude of $\approx 30\%$ (which constraints only very poorly the chirp mass - luminosity distance combination ${\cal M}^{5/3}/D_L$), and the source inclination and polarization angles are recovered at the $\approx 0.3$ rad level. The ongoing Parkes PTA is particularly sensitive to systems located in the southern hemisphere, where at SNR = 10 the source position can be determined with $\Delta \Omega \approx 10\,\mathrm{deg}^2$, but has poorer (by an order or magnitude) performance for sources in the northern hemisphere. 

\end{abstract}

\maketitle

\section{Introduction}
\label{s:intro}

Pulsar Timing Arrays (PTAs), such as the Parkes PTA (PPTA) ~\cite{man08}, the European PTA (EPTA)~\cite{jan08}, Nanograv~\cite{NANOGrav}, the International Pulsar Timing Array (IPTA) project~\cite{HobbsEtAl:2009}, and in the future the Square Kilometre Array (SKA)~\cite{laz09}  provide a unique means to study the population of massive black hole (MBH) binary systems with masses above $\sim 10^7\,M_\odot$ by monitoring stable radio pulsars: in fact,  gravitational waves (GWs) generated by MBH binaries (MBHBs) affect the propagation of electromagnetic signals and leave a distinct signature on the time of arrival of the radio pulses~\cite{EstabrookWahlquist:1975,Sazhin:1978,Detweiler:1979,HellingsDowns:1983}. MBH formation and evolution scenarios~\cite{vhm03, kz06, mal07, yoo07} predict the existence of a large number of MBHBs. Whereas the high redshift, low(er) mass systems will be targeted by the planned Laser Interferometer Space Antenna ({\it LISA}~\cite{bender98})~\cite{enelt,uaiti,ses04,ses05,ses07}, massive and lower redshift ($z\simlt 2$) binaries radiating in the (gravitational) frequency range $\sim 10^{-9}\,\mathrm{Hz} - 10^{-6} \,\mathrm{Hz}$ will be directly accessible to PTAs. These systems imprint a typical signature on the time-of-arrival of radio-pulses at a level of $\approx 1-100$ ns~\cite{papII}, which is comparable with the timing stability of several pulsars~\cite{Hobbs:2009}, with more expected to be discovered and monitored in the future. PTAs therefore provide a direct observational window onto the MBH binary population, and can contribute to address a number of astrophysical open issues, such as the shape of the bright end of the MBH mass function, the nature of the MBH-bulge relation at high masses, and the dynamical evolution at sub-parsec scales of the most massive binaries in the Universe  (particularly relevant to the so-called ``final parsec problem''~\cite{mm03}).  

Gravitational radiation from the cosmic population of MBHBs produces two classes of signals in PTA data: (i) a stochastic GW background generated by the incoherent superposition of radiation from the whole MBHB population~\cite{rr95, phi01, jaffe, jen05, jen06, papI} and (ii) individually resolvable, deterministic signals produced by single sources that are sufficiently massive and/or close so that the gravitational signal stands above the root-mean-square (rms) level of the background ~\cite{papII}. In~\cite{papII} (SVV, hereafter) we explored a comprehensive range of MBH population models and found that, assuming a simple order-of-magnitude criterion to estimate whether sources are resolvable above the background level $\approx 1$-to-10 individual MBHBs could be observed by future PTAs surveys. The observation of GWs from individual systems would open a new avenue for a direct census of the properties of MBHBs, offering invaluable new information about galaxy formation scenarios. The observation of systems at this stage along their merger path would also provide key insights into the understanding of the interaction between MBHBs and the stellar/gaseous environment~\cite{KocsisSesana}, and how these interactions affect the black hole-bulge correlations during the merger process. If an electro-magnetic counterpart of a MBHB identified with PTAs was to be found, such a system could offer a unique laboratory for both accretion physics (on small scales) and the interplay between black holes and their host galaxies (on large scales). 

The prospects of achieving these scientific goals raise the question of what astrophysical information could be extracted from PTA data and the need to quantify the typical statistical errors that will affect the measurements, their dependence on the total number and spatial distribution of pulsars in the array (which affects the surveys observational strategies), and the consequences for multi-band observations. In this paper we estimate the statistical errors that affect the measurements of the source parameters focusing on MBHBs with no spins, in circular orbits, that are sufficiently far from coalescence so that gravitational radiation can be approximated as producing a signal with negligible frequency drift during the course of the observation time, $T \approx 10$ yr ("monochromatic" signal). This is the class of signals that in SVV we estimated to produce the bulk of the observational sample. The extension to eccentric binaries and systems with observable frequency derivative is deferred to a future work. GWs from monochromatic circular binaries constituted by non-spinning MBHs are described by seven independent parameters. We compute the expected statistical errors on the source parameters by evaluating the variance-covariance matrix -- the inverse of the Fisher information matrix -- of the observable parameters. The diagonal elements of such a matrix provide a robust lower limit to the statistical uncertainties (the so-called Cramer-Rao bound~\cite{JaynesBretthorst:2003, Cramer:1946}), which in the limit of high signal-to-noise ratio (SNR) tend to the actual statistical errors. Depending on the actual structure of the signal likelihood function and the SNR this could underestimate the actual errors, see \emph{e.g.}~\cite{NicholsonVecchio:1998,BalasubramanianDhurandhar:1998,Vallisneri:2008} for a discussion in the context of GW observations. Nonetheless, this analysis serves as an important benchmark and can then be refined by carrying out actual analyses on mock data sets and by estimating the full set of (marginalised) posterior density functions of the parameters. The main results of the paper can be summarised as follows:
\begin{itemize} 

\item At least three (not co-aligned) pulsars in the PTA are necessary to fully resolve the source parameters;

\item The statistical errors on the source parameters, at \emph{fixed} SNR, decrease as the number of pulsars in the array increases. The typical accuracy greatly improves by adding pulsars up to $\approx 20$; for larger arrays, the actual gain become progressively smaller because the pulsars ``fill the sky" and the effectiveness of further triangulation saturates. In particular, for a fiducial case of an array of 100 pulsars randomly and uniformly distributed in the sky with optimal coherent SNR = 10 -- which may be appropriate for the SKA --  we find a typical GW source error box in the sky $\approx 40$ deg$^2$ and a fractional amplitude error of $\approx$ 30\%. The inclination and polarization angles can be determined within an error of $\sim 0.3$ rad, and the (constant) frequency is determined to sub-frequency resolution bin accuracy. These results are independent on the source gravitational-wave frequency. 

\item When an anisotropic distribution of pulsars is considered, the typical source sky location accuracy improves linearly with the array sky coverage. The statistical errors on all the other parameters are essentially insensitive to the PTA sky coverage, as long as it covers more than $\sim 1$ srad.

\item The ongoing Parkes PTA aims at monitoring 20 pulsars with a 100 ns timing noise; the targeted pulsars are mainly located in the southern sky. A GW source in that part of the sky could be localized down to a precision of $\lesssim 10$ deg$^2$ at SNR$=10$, whereas in the northern hemisphere, the lack of monitored pulsars limits the error box to $\simgt 200$ deg$^2$. The median of the Parkes PTA angular resolution is  $\approx 130\,(\mathrm{SNR}/10)^{-2}$ deg$^2$.
\end{itemize}
  
The paper is organised as follows. In Section II we describe the GW signal relevant to PTA and we introduce the quantities that come into play in the parameter estimation problem. A review of the Fisher information matrix technique and its application to the PTA case are provided in Section III. Section IV is devoted to the detailed presentation of the results, and in Section V we summarize the main findings of this study and point to future work. Unless otherwise specified, throughout the paper we use geometric units $G=c=1$.

\section{The signal}
\label{s:signal}

Observations of GWs using PTAs exploit the regularity of the time of arrival of radio pulses from pulsars. Gravitational radiation affects the arrival time of the electromagnetic signal by perturbing the null geodesics of photons traveling from a pulsar to the Earth. This was realized over thirty years ago~\cite{EstabrookWahlquist:1975,Sazhin:1978,Detweiler:1979,HellingsDowns:1983}, and the number and timing stability of radio pulsars known today and expected to be monitored with future surveys~\cite{man08,jan08,NANOGrav,laz09} make ensembles of pulsars -- PTAs -- ``cosmic detectors'' of gravitational radiation in the frequency range $\sim 10^{-9}\,\mathrm{Hz} - 10^{-6}\,\mathrm{Hz}$. Here we review the signal produced by a GW source in PTA observations. 
Let us consider a GW metric perturbation $h_{ab}(t)$ in the transverse and traceless gauge (TT) described by the two independent (and time-dependent) polarisation amplitudes $h_+(t)$ and $h_\times(t)$ that carry the information about the GW source. Let us also indicate with $\hat\Omega$ the unit vector that identifies the direction of GW propagation (conversely, the direction to the GW source position in the sky is $-\hat\Omega$). The metric perturbation can therefore be written as:
\be
h_{ab}(t,\hat\Omega) = e_{ab}^+(\hat\Omega) h_+(t,\hat\Omega) + e_{ab}^{\times}(\hat\Omega)\, h_\times(t,\hat\Omega),
\label{e:hab}
\ee
where $e_{ab}^A(\hat\Omega)$ ($A = +\,,\times$) are the polarisation tensors, that are uniquely defined once one specifies the wave principal axes described by the unit vectors $\hm$ and $\hn$ as,
\begin{subequations}
\begin{align}
e_{ab}^+(\ho) &=  \hm_a \hm_b - \hn_a \hn_b\,,
\label{e:e+}
\\
e_{ab}^{\times}(\ho) &= \hm_a \hn_b + \hn_a \hm_b\,.
\label{e:ex}
\end{align}
\end{subequations}
Let us now consider a pulsar emitting radio pulses with a frequency $\nu_0$. Radio waves propagate along the direction described by the unit vector $\hp$, and in the background $h_{ab}$ the frequency of the pulse is affected. For an observer at Earth (or at the Solar System Barycentre), the frequency is shifted according to the characteristic two-pulse function~\cite{EstabrookWahlquist:1975}
\ba
z(t,\ho) & \equiv & \frac{\nu(t) - \nu_0}{\nu_0}
\nonumber\\
& = & \frac{1}{2} \frac{\hp^a\hp^b}{1+\hp^a\ho_a}\Delta h_{ab}(t;\ho)\,.
\label{e:z}
\ea
Here $\nu(t)$ is the received frequency (say, at the Solar System Barycentre), and
\be
\Delta h_{ab}(t) \equiv h_{ab}(t_\mathrm{p},\ho) - h_{ab}(t,\ho)
\label{e:deltah}
\ee
is the difference between the metric perturbation at the pulsar -- with spacetime coordinates $(t_\mathrm{p},\vec{x}_p)$ -- and at the receiver -- with spacetime coordinates $(t,\vec{x})$. The quantity that is actually observed is the time-residual $r(t)$, which is simply the time integral of Eq.~(\ref{e:z}),
\be
r(t) = \int_0^t dt' z(t',\ho)\,.
\label{e:r}
\ee
We can re-write Eq.~(\ref{e:z}) in the form
\be
z(t,\ho) = \sum_A F^A(\ho) \Delta h_{A}(t;\ho)\,,
\label{e:z1}
\ee
where 
\be
F^A(\ho) \equiv \frac{1}{2} \frac{\hp^a\hp^b}{1+\hp^a\ho_a} e_{ab}^A(\ho)
\label{e:FA}
\ee
is the ``antenna beam pattern'', see Eqs~(\ref{e:hab}), (\ref{e:e+}) and~(\ref{e:ex})); here we use the Einstein summation convention for repeated indeces. Using the definitions~(\ref{e:e+}) and~(\ref{e:ex}) for the wave polarisation tensors, it is simple to show that the antenna beam patterns depend on the three direction cosines $\hm \cdot \hp$, $\hn \cdot \hp$ and $\ho \cdot \hp$:
\begin{subequations}
\begin{align}
F^+(\ho) & = \frac{1}{2} \frac{(\hm \cdot \hp)^2 - (\hn \cdot \hp)^2}{1 + \ho \cdot \hp}\,,\\
F^\times(\ho) & = \frac{(\hm \cdot \hp)\,(\hn \cdot \hp)}{1 + \ho \cdot \hp}\,.
\end{align}
\end{subequations}

Let us now consider a reference frame $(x,y,z)$ fixed to the Solar System Barycentre. The source location in the sky is defined by the usual polar angles $(\theta,\phi)$. The unit vectors that define the wave principal axes are given by (cf. Eqs. (B4) and (B5) in Appendix B of ~\cite{Anderson-et-al:2001}; here we adopt the same convention used in high-frequency laser interferometric observations)
\begin{subequations}
\begin{align}
\vec{m} & = 
(\sin\phi \cos\psi - \sin\psi \cos\phi \cos\theta) \hat{x} 
\nonumber\\
& - 
(\cos\phi \cos\psi + \sin\psi \sin\phi \cos\theta) \hat{y}
\nonumber\\
& + 
(\sin\psi \sin\theta) \hat{z}\,,
\label{e:m}\\
\vec{n} & = 
(-\sin\phi \sin\psi - \cos\psi \cos\phi \cos\theta) \hat{x} 
\nonumber\\
& + 
(\cos\phi \sin\psi - \cos\psi \sin\phi \cos\theta) \hat{y} 
\nonumber\\
& + 
(\cos\psi \sin\theta) \hat{z}\,,
\label{e:n}
\end{align}
\end{subequations}
where $\hat{x}$, $\hat{y}$ and $\hat{z}$ are the unit vectors along the axis of the reference frame, $x$, $y$, and $z$, respectively.
The angle $\psi$ is the wave polarisation angle, defined as the angle counter-clockwise about the direction of propagation from the line of nodes to the axis described by $\vec{m}$. The wave propagates in the direction $\hat\Omega = \vec{m} \times \vec{n}$, which is explicitly given by:
\be
\ho = 
- (\sin\theta \cos\phi)\, \hat{x}
- (\sin\theta \sin\phi)\, \hat{y}
- \cos\theta \hat{z}\,.
\ee
Analogously, the unit vector
\be
\hp_\alpha = 
 (\sin\theta_\alpha \cos\phi_\alpha)\, \hat{x}
 + (\sin\theta_\alpha \sin\phi_\alpha)\, \hat{y}
 + \cos\theta_\alpha \hat{z}
\ee
identifies the position in the sky of the $\alpha$-th pulsar using the polar angles $(\theta_\alpha,\phi_\alpha)$. 

We will now derive the expression of the PTA signal, Eq.~\ref{e:z}, produced by a circular, non-precessing binary system of MBHs emitting almost monochromatic radiation, \emph{i.e} with negligible frequency drift during the observation time, $T\approx 10$ yr. The results are presented in Section~\ref{ss:timing-residuals}. In the next sub-section, we firstly justify the astrophysical assumptions.

\subsection{Astrophysical assumptions}
\label{ss:astrophysics}

Let us justify (and discuss the limitations of) the assumptions that we have made on the nature of the sources, that lead us to consider circular, non-precessing binary systems generating quasi-monochromatic radiation, before providing the result in Eq.~(\ref{researth}). We derive general expressions for the phase evolution displacement introduced by the frequency drift and by the eccentricity induced periastron precession, and the change in the orbital angular momentum direction caused by the spin-orbit coupling induced precession. The size of each of these effects is then evaluated by considering a realistic (within our current astrophysical understanding) selected population of resolvable MBHBs taken from SVV. Throughout the paper we will consider binary systems with masses $m_1$ and $m_2$ ($m_2 \le m_1$), and \emph{chirp mass} ${\cal M}=m_1^{3/5}m_2^{3/5}/(m_1+m_2)^{1/5}$, emitting a GW frequency $f$. We also define $M = m_1 + m_2$, $\mu = m_1 m_2/m$ and $q=m_2/m_1$, the total mass, the reduced mass and the mass ratio, respectively. Our notation is such that all the quantities are the observed (redshifted) ones, such that \emph{e.g.} the {\it intrinsic} (rest-frame) mass of the primary MBH is $m_{1,r}=m_1/(1+z)$ and the {\it rest frame} GW frequency is $f_r=f(1+z)$. We normalize all the results to
\begin{align}
& M_9 = \frac{M}{10^9\,\msun}\,,
\nonumber\\
& {\cal M}_{8.5} = \frac{{\cal M}}{10^{8.5}\,\msun}\,,
\nonumber\\
& f_{50} = \frac{f}{50\,{\rm nHz}}\,,
\nonumber\\
& T_{10} = \frac{T}{10\,{\rm yr}}\,,
\nonumber
\end{align}
which are the typical values for individually resolvable sources found in SVV, and the typical observation timespan.

\subsubsection{Gravitational wave frequency evolution}

A binary with the properties defined above, evolves due to radiation reaction through an adiabatic in-spiral phase, with \emph{GW frequency} $f(t)$ changing at a rate (at the leading Newtonian order)
\be
\frac{df}{dt} = \frac{96}{5}\pi^{8/3} {\cal M}^{5/3} f^{11/3}\,.
\label{e:dfdt}
\ee
The in-spiral phase terminates at the last stable orbit (LSO), that for a Schwarzschild black hole in circular orbit corresponds to the frequency
\be
f_\mathrm{LSO} = 4.4\times 10^{-6}\,M_9^{-1}\,\,\mathrm{Hz}\,.
\label{e:flso}
\ee
The observational window of PTAs is set at low frequency by the overall duration of the monitoring of pulsars $T \approx 10$ yr, and at high frequency by the cadence of the observation, $\approx 1$ week: the PTA observational window is therefore in the range $\sim 10^{-9} - 10^{-6}$ Hz. In SVV we explored the physical properties of MBHBs that are likely to be observed in this frequency range: PTAs will resolve binaries with $m_{1,2} \simgt 10^8 M_\odot$ and in the frequency range $\approx 10^{-8} - 10^{-7}$ Hz.  In this mass-frequency region, PTAs will observe the in-spiral portion of the coalescence of a binary system and one can ignore post-Newtonian corrections to the amplitude and phase evolution, as the velocity of the binary is:
\begin{align}
v & = (\pi f M)^{2/3}\,,
\nonumber\\
&  = 1.73\times 10^{-2} M_9^{2/3}f_{50}^{2/3}\,.
\label{e:v}
\end{align}
Stated in different terms, the systems will be far from plunge, as the time to coalescence for a binary radiating at frequency $f$ is (at the leading Newtonian quadrupole order and for a circular orbit system)
\be
t_\mathrm{coal} \simeq 4\times 10^3\,{\cal M}_{8.5}^{-5/3}\,f_{50}^{-8/3}\,\mathrm{yr}\,.
\label{e:tcoal}
\ee
As a consequence the frequency evolution during the observation time is going to be small, and can be neglected. In fact, it is simple to estimate the total frequency shift of radiation over the observation period 
\be
\Delta f \approx \dot{f} T \approx 0.05\,{\cal M}_{8.5}^{5/3}\,f_{50}^{11/3}\,T_{10}\,\,\, \mathrm{nHz}\,,
\label{e:fdrift}
\ee
which is negligible with respect to the frequency resolution bin $\approx 3 T_{10}^{-1}$ nHz; correspondingly, the additional phase contribution
\be
\Delta \Phi  \approx \pi \dot{f} T^2 \approx 0.04\,{\cal M}_{8.5}^{5/3}\,f_{50}^{11/3}\,T_{10}^2\,\,\, \mathrm{rad},
\label{e:phasedrift}
\ee
is much smaller than 1 rad. Eqs.~(\ref{e:fdrift}) and~(\ref{e:phasedrift}) clearly show that it is more than legitimate in this initial study to ignore any frequency derivative, and treat gravitational radiation as \emph{monochromatic} over the observational period. 

\subsubsection{Spin effects}

We now justify our assumption of neglecting the spins in the modelling of the waveform. From an astrophysical point of view, very little precise information about the spin of MBHs can be extracted directly from observations. However, several theoretical arguments support the existence of a population of rapidly spinning MBHs. If coherent accretion from a thin disk \cite{ss73} is the dominant growth mechanism, then MBH spin-up is inevitable \cite{thorne74}; jet production in active galactic nuclei is best explained by the presence of rapidly spinning MBHs \cite{nemmen07}; in the hierarchical formation context, though MBHB mergers tend to spin down the remnant \cite{hughes03}, detailed growth models that take into account both mergers and accretion lead to populations of rapidly spinning MBHs \cite{vp05,bv08}. Spins have two main effects on the gravitational waveforms emitted during the in-spiral: (i) they affect the phase evolution~\cite{BlanchetEtAl:1995}, and (ii) they cause the orbital plane to precess through spin-orbit and spin-spin coupling~\cite{ApostolatosEtAl:1994,Kidder:1995}. The effect of the spins on the phase evolution is completely negligible for the astrophysical systems observable by PTAs: the additional phase contribution enters at the lowest order at the post$^{1.5}$-Newtonian order, that is proportional to $v^3$, and we have already shown that $v \ll 1$, see Eq.~(\ref{e:v}). Precession would provide a characteristic imprint on the signal through amplitude and phase modulations produced by the orbital plane precession, and as a consequence the time-dependent polarisation of the waves as observed by a PTA. It is fairly simple to quantify the change of the orientation of the orbital angular momentum unit vector $\vec{L}$ during a typical observation. The rate of change of the precession angle is at the leading order:
\be
\frac{d\alpha_p}{dt} = \left(2 + \frac{3 m_2}{2 m_1}\right) \frac{L + S}{a^3}
\label{e:dalphadt}
\ee
where $L = \sqrt{a \mu^2 M}$ is the magnitude of the orbital angular momentum and $S$ is the total intrinsic spin of the black holes. As long as $\mu/M\gg v/c$, we have that $L \gg S$. This is always the case for resolvable MBHBs; we find indeed that these systems are in general characterised by $q\gtrsim0.1$ (therefore $\mu/M\gtrsim0.1$), while from Eq. (\ref{e:v}) we know that in general $v/c\sim0.01$ .  In this case, from Eq.~(\ref{e:dalphadt}) one obtains
\ba
\Delta \alpha_p & \approx&  2\pi^{5/3} \left(1 + \frac{3 m_2}{4 m_1}\right) \mu M^{-1/3} f^{5/3} T
\nonumber\\
& \approx & 0.8 \left(1 + \frac{3 m_2}{4 m_1}\right)\left(\frac{\mu}{M}\right)M_9^{2/3}f_{50}^{5/3}T_{10}\,\mathrm{rad}\,, 
\label{spin}
\ea
which is independent of $S$. The effect is maximum for equal mass binaries, $m_1 = m_2$, ${\mu}/{M} = 0.25$; in this case $\Delta \alpha_p \approx 0.3$ rad. It is therefore clear that in general spins will not play an important role, and we will neglect their effect in the modeling of signals at the PTA output. It is however interesting to notice that for a $10^9 M_\odot$ binary system observed for 10 years at $\approx 10^{-7}$ Hz, which is consistent with astrophysical expectations (see SVV) the orientation of the orbital angular momentum would change by $\Delta \alpha_p \approx 1$ rad. The Square-Kilometre-Array has therefore a concrete chance of detecting this signature, and to provide direct insights onto MBH spins.

\subsubsection{Eccentricity of the binary}

Let us finally consider the assumption of circular orbits, and the possible effects of neglecting eccentricity in the analysis. The presence of a residual eccentricity at orbital separations corresponding to the PTA observational window has two consequences on the observed signal: (i) the power of radiation is not confined to the harmonic at twice the orbital frequency but is spread on the (in principle infinite) set of harmonics at integer multiples of the inverse of the orbital period, and (ii) the source periapse precesses in the plane of the orbit at a rate 
\begin{align}
\frac{d\gamma}{dt} & = 3\pi f \frac{\left(\pi f M\right)^{2/3}}{\left(1 - e^2\right)}\,,
\nonumber\\
& \simeq 3.9\times10^{-9} \left(1 - e^2\right)^{-1}\,M_{9}^{2/3}\,f_{50}^{5/3}\,\mathrm{rad\,\,s}^{-1}
\label{e:dgammadt}
\end{align}
which introduces additional modulations in phase and (as a consequence) amplitude in the signal recorded at the Earth. In Eq.~(\ref{e:dgammadt}) $\gamma(t)$ is the angle of the periapse measured with respect to a fixed frame attached to the source. We now briefly consider the two effects in turn. The presence of eccentricity "splits" each polarisation amplitude $h_+(t)$ and $h_\times(t)$ into harmonics according to (see \emph{e.g.} Eqs.~(5-6) in Ref.~\cite{WillemsVecchioKalogera:2008} and references therein):
\ba
h^{+}_n(t) & = & A \Bigl\{-(1 + \cos^2\iota)u_n(e) \cos\left[\frac{n}{2}\,\Phi(t) + 2 \gamma(t)\right] 
\nonumber \\ 
& & -(1 + \cos^2\iota) v_n(e) \cos\left[\frac{n}{2}\,\Phi(t) - 2 \gamma(t)\right]
\nonumber \\
 & & + \sin^2\iota\, w_n(e) \cos\left[\frac{n}{2}\,\Phi(t)\right] \Bigr\},
\label{e:h+}\\
h^{\times}_{n}(t) & = & 2 A \cos\iota \Bigl\{u_n(e) \sin\left[\frac{n}{2}\,\Phi(t) + 2 \gamma(t)\right] 
\nonumber\\
& & + v_n(e) \sin(\left[\frac{n}{2}\,\Phi(t) - 2 \gamma(t)\right]) \Bigr\}\,,
\label{e:hx}
\ea
where
\be
\Phi(t) = 2\pi\int^t f(t') dt'\,,
\label{e:Phi}
\ee
is the GW phase and $f(t)$ the instantaneous GW frequency corresponding to twice the inverse of the orbital period. The source inclination angle $\iota$ is defined as $\cos\iota = -\ho^a {\hat L}_a$, where ${\hat L}$ is the unit vector that describes the orientation of the source orbital plane, and the amplitude coefficients $u_n(e)$,  $v_n(e)$, and $w_n(e)$ are linear combinations of the Bessel functions of the first kind $J_{n}(ne)$, $J_{n\pm 1}(ne)$ and $J_{n\pm 2}(ne)$. For an astrophysically plausible range of eccentricities $e\simlt 0.3$ -- see Fig.~\ref{fig1a} and the discussion below -- $|u_n(e)| \gg |v_n(e)|\,,|w_n(e)|$ and most of the power will still be confined into the $n=2$ harmonic at twice the orbital frequency, see \emph{e.g.} Fig. 3 of\ Ref.~\cite{PetersMathews:1963}. On the other hand, the change of the periapse position even for low eccentricity values may introduce significant phase shifts over coherent observations lasting several years. In fact the phase of the recorded signal is shifted by an additional contribution $2\gamma(t)$. This means that the actual frequency of the observed signal recorded at the instrument corresponds to $f(t) + {\dot{\gamma}}/{\pi}$ and differs by a measurable amount from $f(t)$. Nonetheless, one can still model the radiation observed at the PTA output as monochromatic, as long as th periapse precession term ${\dot{\gamma}}/{\pi}$ introduces a phase shift $\Delta \Phi_\gamma $ quadratic in time that is $\ll 1$ rad, which is equivalent to the condition that we have imposed on the change of the phase produced by the frequency shift induced by radiation reaction, see Eqs.~(\ref{e:fdrift}) and~(\ref{e:phasedrift}). From Eq.~(\ref{e:dgammadt}) and~(\ref{e:dfdt}), this condition yields:  
\begin{align}
\Delta \Phi_\gamma & \approx \frac{d^2\gamma}{dt^2} T^2 = \frac{96\pi^{13/3}}{\left(1 - e^2\right)} M^{2/3}{\cal M}^{5/3} f^{13/3} T^2
\nonumber\\
& \approx 2\times10^{-3} \left(1 - e^2\right)^{-1} M_9^{2/3}{\cal M}_{8.5}^{5/3}\,f_{50}^{13/3}\,T_{10}^2\,\mathrm{rad}\,.
\label{e:dgamma}
\end{align}
We therefore see that the effect of the eccentricity will be in general negligible.

\subsubsection{Tests on a massive black hole population}

\begin{figure}
\centerline{\psfig{file=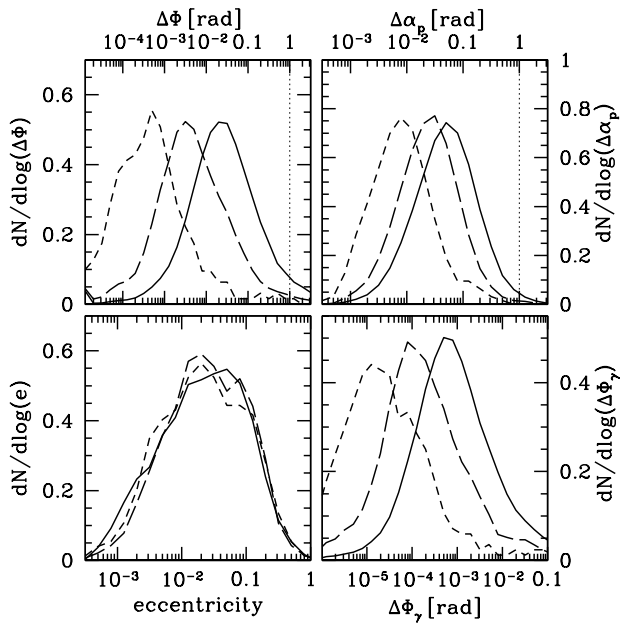,width=84.0mm}}
\caption{Testing the circular monochromatic non--spinning binary approximation. Upper left panel: distribution of the phase displacement $\Delta \Phi$ introduced by the frequency drift of the binaries. Upper right panel: change in the orbital angular momentum direction $\Delta \alpha_p$ introduced by the spin-orbit coupling. Lower left panel: eccentricity distribution of the systems. Lower right panel: distribution of phase displacement $\Delta \Phi_\gamma$ induced by relativistic periastron precession due to non-zero eccentricity of the binaries. The distributions are constructed considering all the resolvable MBHBs with residuals $>1$ns (solid lines), 10ns (long--dashed lines) and 100ns (short--dashed lines), found in 1000 Monte Carlo realizations of the Tu-SA models described in SVV, and they are normalised so that their integrals are unity.}
\label{fig1a}
\end{figure}

We can quantify more rigorously whether the assumption of monochromatic signal at the PTA output is justified, by evaluating the distributions of $\Delta \Phi$, $\Delta \alpha_p$ and $\Delta \Phi_\gamma$ on an astrophysically motivated population of resolvable MBHBs. We consider the Tu-SA MBHB population model discussed in SVV (see Section 2.2 of SVV for a detailed description) and we explore the orbital evolution, including a possible non-zero eccentricity of the observable systems. The binaries are assumed to be in circular orbit at the moment of pairing and are self consistently evolved taking into account stellar scattering and GW emission~\cite{Sesana-prep}. We generate 1000 Monte Carlo realisations of the entire population of GW signals in the PTA band and we collect the individually resolvable sources generating coherent timing residuals greater than 1, 10 and 100 ns, respectively, over 10 years. In Fig. \ref{fig1a} we plot the distributions relevant to this analysis. We see from the two upper panels, that in general, treating the system as "monochromatic" with negligible spin effects is a good approximation. If we consider a 1 ns threshold (solid lines), the phase displacement $\Delta \Phi$ introduced by the frequency drift and the orbital angular momentum direction change $\Delta \alpha_p$ due two spin-orbit coupling are always $<1$ rad, and in $\sim 80$\% of the cases are $<0.1$ rad. The lower left panel of Fig. \ref{fig1a} shows the eccentricity distribution of the same sample of individually resolvable sources. Almost all the sources are characterised by $e \simlt 0.1$ with a long tail extending down to $e \simlt 10^{-3}$ in the PTA band. The typical periastron precession--induced additional phase $2\dot{\gamma}T$ can be larger than 1 rad. However, this additional contribution grows linearly with time, and, as discussed before, will result in a measured frequency which differs from the intrinsic one by a small amount $\dot{\gamma}/\pi\lesssim 1$nHz. The ``non-monocromatic'' phase contribution $\Delta \Phi_\gamma$ that changes quadratically with time and is described by Eq. (\ref{e:dgamma}) is instead plotted in the lower right panel of Fig. \ref{fig1a}. Values of $\Delta \Phi_\gamma$ are typically of the order $10^{-3}$, completely negligible in the context of our analysis. Note that, as a general trend, increasing the threshold in the source--induced timing residuals to 10 and 100 ns, all the effects tend to be suppressed. This is because resolvable sources generating larger residuals are usually found at lower frequencies, and all the effects have a steep dependence on frequency -- see Eqs.~(\ref{e:phasedrift}),~(\ref{spin}) and~(\ref{e:dgamma}). This means that none of the effects considered above should be an issue for ongoing PTA campaigns, which aim to reach a total sensitivity of $\gtrsim30$ ns, but may possibly play a role in recovering sources at the level of a few ns, which is relevant for the planned SKA. Needless to say that a residual eccentricity at the time of pairing may result in larger values of $e$ than those shown in Fig.~\ref{fig1a}~\cite{Sesana-prep}, causing a significant scatter of the signal power among several different harmonics; however, the presence of gas may lead to circularization before they reach a frequency $\approx 10^{-9}$ Hz (see, e.g.,~\cite{dot07}). Unfortunately, little is known about the eccentricity of subparsec massive binaries, and here we tackle the case of circular systems, deferring the study of precessing eccentric binaries to future work.

\subsection{Timing residuals}
\label{ss:timing-residuals}

\begin{figure}
\centerline{\psfig{file=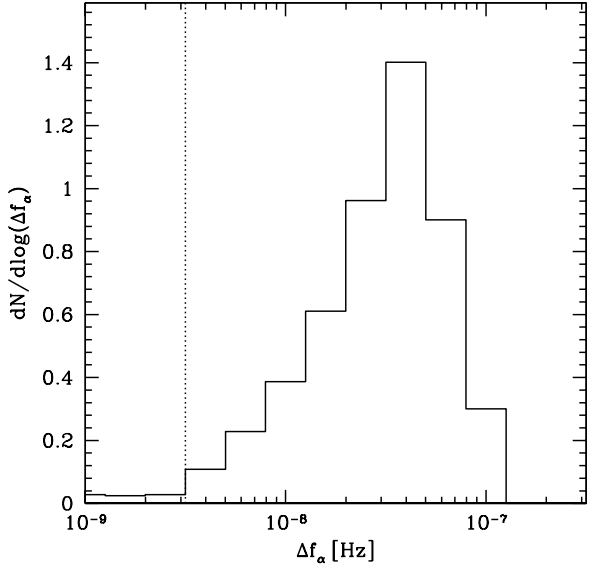,width=84.0mm}}
\caption{Normalized distribution of $\Delta f_{\alpha}$ (see text) for the same sample of MBHBs considered in Fig. \ref{fig1a}, assuming observations with 100 isotropically distributed pulsars in the sky at a distance of 1 kpc. The vertical dotted line marks the width of the array's frequency resolution bin $\Delta f_r=1/T$ ($\approx 3\times 10^{-9}$Hz for $T=10$yr).}
\label{fig1b}
\end{figure}

We have shown that the assumption of circular, monochromatic, non-precessing binary is astrophysically reasonable, surely for this initial exploratory study. We now specify the signal observed at the output, Eq.~(\ref{e:r}), in this approximation. The two independent polarisation amplitudes generated by a binary system, Eqs.~(\ref{e:h+}) and~(\ref{e:hx}), can be written as:
\begin{subequations}
\begin{align}
h_+(t) & = A_\mathrm{gw} a(\iota) \cos\Phi(t)\,,
\label{e:h+1}
\\
h_{\times}(t) &= A_\mathrm{gw} b(\iota) \sin\Phi(t)\,,
\label{e:hx1}
\end{align}
\end{subequations}
where
\be
A_\mathrm{gw}(f) = 2 \frac{{\cal M}^{5/3}}{D}\,\left[\pi f(t)\right]^{2/3}
\label{e:Agw}
\ee
is the GW amplitude, $D$ the luminosity distance to the GW source, $\Phi(t)$ is the GW phase given by Eq. (\ref{e:Phi}), and $f(t)$ the instantaneous GW frequency (twice the inverse of the orbital period). The two functions
\begin{subequations}
\begin{align}
a(\iota) & = 1 + \cos^2 \iota
\label{e:aiota}
\\
b(\iota) &= -2 \cos\iota
\label{e:biota}
\end{align}
\end{subequations}
depend on the source inclination angle $\iota$, defined in the previous Section.

As described in Section II, Eqs.~(\ref{e:deltah}) and~(\ref{e:r}), the response function of each individual pulsar $\alpha$ consists of two terms, namely, the perturbation registered at the Earth at the time $t$ of data collection ($h_{ab}(t,\ho)$), and the perturbation registered at the pulsar at a time $t-\tau_\alpha$ ($h_{ab}(t-\tau_\alpha,\ho)$), where $\tau_\alpha$ is the light-travel-time from the pulsar to the Earth given by:
\ba
\tau_\alpha & = & L_\alpha (1 + \ho \cdot \hp_\alpha)
\nonumber\\
& \simeq & 1.1\times 10^{11}\,\frac{L_\alpha}{1\,\mathrm{kpc}}\,(1 + \ho \cdot \hp_\alpha)\,\mathrm{s},
\ea
where $L_\alpha$ is the distance to the pulsar. We can therefore formally write the observed timing residuals, Eq.~(\ref{e:r}) for each pulsar $\alpha$ as:
\be
r_\alpha(t) = r_\alpha^{(P)}(t) +  r_\alpha^{(E)}(t)\,,
\label{e:r1}
\ee
where $P$ and $E$ label the ``pulsar'' and ``Earth'' contribution, respectively. During the time $\tau_\alpha$ the frequency of the source -- although "monochromatic" over the time of observation $T$ of several years -- changes by
\be
\Delta f_{\alpha}=\int_{t-\tau_\alpha}^{t} \frac{df}{dt} dt \sim \frac{df}{dt}\tau_\alpha \approx 15\,{\cal M}_{8.5}^{5/3}f_{50}^{11/3}\tau_{\alpha,1}\,\,\,\mathrm{nHz},
\ee
where $\tau_{\alpha,1}$ is the pulsar-Earth light-travel-time normalized to a distance of 1 kpc. The frequency shift $\Delta f_{\alpha}$ depends both on the parameters of the source (emission frequency and chirp mass) and the properties of the pulsar (distance and sky location with respect to the source).  We can quantify this effect over an astrophysically plausible sample of GW sources by considering the population shown in Fig.~\ref{fig1a}. Let us consider the same set of resolvable sources as above, and assume detection with a PTA of 100 pulsars randomly distributed in the sky, but all at a distance of 1 kpc. For each source we consider all the $\Delta f_{\alpha}$ related to each pulsar and we plot the results in Fig. \ref{fig1b}. The distribution has a peak around $\sim5\times 10^{-8}$ Hz, which is $\sim 10$ times larger than the typical frequency resolution bin for an observing time $T\approx 10$ yr. This means that \emph{the signal associated to each pulsar generates at the PTA output two monochromatic terms at two distinct frequencies.} All the "Earth-terms"corresponding to each individual pulsar share the same frequency and phase. They can therefore be coherently summed among the array, building up a distinct monochromatic peak which is not going to be affect by the pulsar terms (also known as  "self-noise") which usually happen to be at much lower frequencies. The contribution to the Earth term from each individual pulsar can be written as 
\ba
r_\alpha^{(E)}(t) & = & R \,[a\, F^+_\alpha\,(\sin\Phi(t)-\sin\Phi_0) 
\nonumber\\
& - & b\, F^\times_\alpha(\cos\Phi(t)-\cos\Phi_0)\,] ,
\label{researth}
\ea
with 
\be
R=\frac{A_{\rm gw}}{2\pi f}
\label{erre}
\ee
and $\Phi(t)$ given by Eq. (\ref{e:Phi}). The Earth timing residuals are therefore described by a 7-dimensional vector encoding all (and only) the parameters of the source:
\be
\vec{\lambda} = \{R,\theta,\phi,\psi,\iota,f,\Phi_0\}\,.
\label{par}
\ee
Conversely, each individual pulsar term is characterized by a different amplitude, frequency and phase, that crucially \emph{depend also on the poorly constrained distance $L_\alpha$ to the pulsar}. In order to take advantage of the power contained in the pulsar term, one needs to introduce an additional parameter for each pulsar in the PTA. As a consequence, this turns a 7-parameter reconstruction problem into a $7+M$ parameter problem. More details about the PTA response to GWs are given in Appendix A. In this paper, we decided to consider simply the Earth-term (at the expense of a modest loss in total SNR) given by Eq. (\ref{researth}), which is completely specified by the 7-parameter vector~(\ref{par}). At present, it is not clear whether it would also be advantageous to include into the analysis the pulsar-terms, that require the addition of $M$ unknown search parameters. This is an open issue that deserves further investigations and will be considered in a future paper.

\section{Parameter estimation}
\label{s:fim}

In this section we briefly review the basic theory and key equations regarding the estimate of the statistical errors that affect the measurements of the source parameters. For a comprehensive discussion of this topic we refer the reader to~\cite{JaynesBretthorst:2003}.

The whole data set collected using a PTA consisting of $M$ pulsars can be schematically represented as a vector
\be
\vec{d} = \left\{d_1, d_2, \dots, d_M\right\}\,,
\label{e:vecd}
\ee
where the data form the monitoring each pulsar $(\alpha = 1,\dots,M)$ are given by
\be
d_\alpha(t) = n_\alpha(t) + r_\alpha(t;\vec{\lambda})\,.
\label{e:da}
\ee
In the previous equation $r_\alpha(t;\vec{\lambda})$, given by Eq.~(\ref{researth}), is the GW contribution to the timing residuals of the $\alpha$-th pulsar (the signal) -- to simplify notation we have dropped (and will do so from now on) the index "E", but it should be understood as we have stressed in the previous section that we will consider only the Earth-term in the analysis -- and  $n_\alpha(t)$ is the noise that affects the observations. For this analysis we make the usual (simplifying) assumption that $n_\alpha$ is a zero-mean Gaussian and stationary random process characterised by the one-sided power spectral density $S_\alpha(f)$.

The inference process in which we are interested in this paper is how well one can infer the actual value of the unknown parameter vector $\vec\lambda$, Eq.~(\ref{par}), based on the data $\vec{d}$, Eq.~(\ref{e:vecd}), and any prior information on $\vec\lambda$ available before the experiment. Within the Bayesian framework, see \emph{e.g.}~\cite{bayesian-data-analysis}, one is therefore interested in deriving the posterior probability density function (PDF) $p(\vec\lambda | \vec d)$ of the unknown parameter vector given the data set and the prior information. Bayes' theorem yields
\be
p(\vec\lambda | \vec d) = \frac{p(\vec\lambda)\,p(\vec d|\vec\lambda)}{p(\vec d)}\,,
\label{e:posterior}
\ee
where $p(\vec d|\vec\lambda)$ is the likelihood function, $p(\vec\lambda)$ is the prior probability density of $\vec\lambda$, and $p(\vec d)$ is the marginal likelihood or evidence. In the neighborhood of the maximum-likelihood estimate value $\hat{{\vec \lambda}}$, the likelihood function can be approximated as a multi-variate Gaussian distribution,
\be
p(\vec\lambda | \vec d) \propto p(\vec\lambda)
 \exp{\left[-\frac{1}{2}\Gamma_{ab} \Delta\lambda_a \Delta\lambda_b\right]}\,,
 \ee
where $ \Delta\lambda_a = \hat{\lambda}_a - {\lambda}_a$ and the matrix $\Gamma_{ab}$ is the Fisher information matrix; here the indexes $a,b = 1,\dots, 7$ label the components of $\vec{\lambda}$.. Note that we have used Einstein's summation convention (and we do not distinguish between covariant and contravariant indeces). In the limit of large SNR, $\hat{{\vec \lambda}}$ tends to ${{\vec \lambda}}$, and the inverse of the Fisher information matrix provides a lower limit to the error covariance of unbiased estimators of ${{\vec \lambda}}$, the so-called Cramer-Rao bound~\cite{Cramer:1946}.  The variance-covariance matrix is simply the inverse of the Fisher information matrix, and its elements are
\begin{subequations}
\ba
\sigma_a^2 & =  & \left(\Gamma^{-1}\right)_{aa}\,,
\label{e:sigma}
\\
c_{ab} & = & \frac{\left(\Gamma^{-1}\right)_{ab}}{\sqrt{\sigma_a^2\sigma_b^2}}\,,
\label{e:cab}
\ea
\end{subequations}
where $-1\le c_{ab} \le +1$ ($\forall a,b$) are the correlation coefficients. We can therefore interpret $\sigma_a^2$ as a way to quantifying the expected uncertainties on the measurements of the source parameters. We refer the reader to~\cite{Vallisneri:2008} and references therein for an in-depth discussion of the interpretation of the inverse of the Fisher information matrix in the context of assessing the prospect of the estimation of the source parameters for GW observations. Here it suffices to point out that MBHBs will likely be observed at the detection threshold (see SVV), and the results presented in Section~\ref{s:results} should indeed be regarded as lower-limits to the statistical errors that one can expect to obtain in real observations, see \emph{e.g.}~\cite{NicholsonVecchio:1998,BalasubramanianDhurandhar:1998,Vallisneri:2008}.

One of the parameters that is of particular interest is the source sky location, and we will discuss in the next Section the ability of PTA to define an error box in the sky. Following Ref.~\cite{Cutler:1998}, we define the PTA angular resolution, or source error box as
\be
\Delta \Omega=2\pi\sqrt{({\rm sin}\theta \Delta \theta \Delta \phi)^2-({\rm sin}\theta c^{\theta\phi})^2}\,;
\label{domega}
\ee
with this definition, the probability for a source to lay \emph{outside} the solid angle $\Delta \Omega_0$ is $e^{-\Delta \Omega_0/\Delta \Omega}$~\cite{Cutler:1998}.

We turn now on the actual computation of the Fisher information matrix $\Gamma_{ab}$. First of all we note that in observations of multiple pulsars in the array, one can safely consider the data from different pulsars as independent, and the likelihood function of $\vec{d}$ is therefore
\ba
p(\vec d|\vec\lambda) & = & \prod_\alpha p(d_\alpha|\vec\lambda)
\nonumber\\
& \propto &  \exp{\left[-\frac{1}{2}\Gamma_{ab} \Delta\lambda_a \Delta\lambda_b\right]}\,,
\ea
where the Fisher information matrix that characterises the \emph{joint} observations in the equation above is simply given by
\be
\Gamma_{ab} = \sum_\alpha \Gamma_{ab}^{(\alpha)}\,.
\ee
$\Gamma_{ab}^{(\alpha)}$ is the Fisher information matrix relevant to the observation with the $\alpha-$th pulsar, and is simply related to the derivatives of the GW signal with respect to the unknown parameters integrated over the observation:
\be
\Gamma_{ab}^{(\alpha)} = \left(\frac{\partial r_\alpha(t; \vec\lambda)}{\partial\lambda_a} \Biggl|\Biggr.\frac{\partial r_\alpha(t; \vec\lambda)}{\partial\lambda_b}
\right)\,,
\label{e:Gamma_ab_a}
\ee
where the inner product between two functions $x(t)$ and $y(t)$ is defined as
\begin{subequations}
\ba
(x|y) & = & 2 \int_{0}^{\infty} \frac{\tilde x^*(f) \tilde y(f) +  \tilde x(f) \tilde y^*(f)}{S_n(f)} df\,,
\label{e:innerxy}
\\
& \simeq & \frac{2}{S_0}\int_0^{T} x(t) y(t) dt\,,
\label{e:innerxyapprox}
\ea
\end{subequations}
and
\be
\tilde x(f) = \int_{-\infty}^{+\infty} x(t) e^{-2\pi i f t}
\label{e:tildex}
\ee
is the Fourier Transform of a generic function $x(t)$. The second equality, Eq.~(\ref{e:innerxyapprox}) is correct only in the case in which the noise spectral density is approximately constant (with value $S_0$) across the frequency region that provides support for the two functions $\tilde x(f)$ and $\tilde y(f)$. Eq.~(\ref{e:innerxyapprox}) is appropriate to compute the scalar product for the observation of gravitational radiation from MBHBs whose frequency evolution is negligible during the observation time, which is astrophysically justified as we have shown in Section~\ref{s:intro}.

In terms of the inner product $(.|.)$ -- Eqs.~(\ref{e:innerxy}) and~(\ref{e:innerxyapprox}) -- the optimal SNR at which a signal can be observed using $\alpha$ pulsars is
\be
{\rm SNR}_\alpha^2 = (r_\alpha | r_\alpha)\,,
\label{e:rhoalpha}
\ee
and the total coherent SNR produced by timing an array of $M$ pulsars is:
\be
{\rm SNR}^2 = \sum_{\alpha = 1}^M {\rm SNR}_\alpha^2\,.
\label{e:rho}
\ee

\section{Results}
\label{s:results}

\begin{table*}
\begin{center}
\begin{tabular}{ll|cccccc}
\hline
$M$ $\,\,$& $\Delta \Omega_\mathrm{PTA} [{\rm srad}]$ $\,\,$& $\,\,\Delta\Omega$ [deg$^2$] $\,\,$& $\,\,\Delta R/R$ $\,\,$& $\,\,\Delta \iota$ [rad] $\,\,$& $\,\,\Delta \psi$ [rad] $\,\,$& $\,\,\Delta f/(10^{-10}{\rm Hz})$ $\,\,$& $\,\,\Delta \Phi_0$ [rad] $\,\,$\\
\hline
3 & $4\pi$ & $2858^{+5182}_{-1693}$ & $2.00^{+4.46}_{-1.21}$ & $1.29^{+5.02}_{-0.92}$ & $2.45^{+9.85}_{-1.67}$ & $1.78^{+0.46}_{0.40}$ & $3.02^{+16.08}_{-2.23}$\\
4 & $4\pi$ & $804^{+662}_{-370}$ & $0.76^{+1.19}_{-0.39}$ & $0.55^{+1.79}_{-0.36}$ & $0.89^{+2.90}_{-0.54}$ & $1.78^{+0.41}_{-0.33}$ & $1.29^{+5.79}_{-0.88}$\\
5 & $4\pi$ & $495^{+308}_{-216}$ & $0.54^{+0.84}_{-0.25}$ & $0.43^{+1.35}_{-0.28}$ & $0.65^{+2.10}_{-0.39}$ & $1.78^{+0.36}_{-0.30}$ & $0.98^{+4.27}_{-0.62}$\\
10 & $4\pi$ & $193^{+127}_{-92}$ & $0.36^{+0.57}_{-0.17}$ & $0.30^{+0.93}_{-0.19}$ & $0.42^{+1.49}_{-0.25}$ & $1.78^{+0.26}_{-0.23}$ & $0.71^{+3.01}_{-0.41}$\\
20 & $4\pi$ & $99.1^{+65.3}_{-44.6}$ & $0.31^{+0.51}_{-0.15}$ & $0.27^{+0.83}_{-0.16}$ & $0.35^{+1.34}_{-0.21}$ & $1.78^{+0.22}_{-0.20}$ & $0.65^{+2.66}_{-0.36}$\\
50 & $4\pi$ & $55.8^{30.5+}_{-23.0}$ & $0.30^{+0.49}_{-0.14}$ & $0.25^{+0.80}_{-0.15}$ & $0.31^{+1.26}_{-0.19}$ & $1.78^{+0.17}_{-0.16}$ & $0.60^{+2.56}_{-0.33}$\\
100 & $4\pi$ & $41.3^{+18.4}_{-15.3}$ & $0.29^{+0.48}_{-0.14}$ & $0.25^{+0.77}_{-0.15}$ & $0.31^{+1.24}_{-0.19}$ & $1.78^{+0.13}_{-0.12}$ & $0.60^{+2.49}_{-0.33}$\\
200 & $4\pi$ & $32.8^{+13.5}_{-11.1}$ & $0.29^{+0.48}_{-0.14}$ & $0.24^{+0.75}_{-0.15}$ & $0.29^{+1.21}_{-0.18}$ & $1.78^{+0.13}_{-0.12}$ & $0.59^{+2.50}_{-0.31}$\\
500 & $4\pi$ & $26.7^{+8.4}_{-8.2}$ & $0.29^{+0.48}_{-0.14}$ & $0.24^{+0.75}_{-0.15}$ & $0.29^{+1.21}_{-0.18}$ & $1.78^{+0.08}_{-0.08}$ & $0.59^{+2.50}_{-0.31}$\\
1000 & $4\pi$ & $23.2^{+6.7}_{-6.8}$ & $0.29^{+0.48}_{-0.14}$ & $0.24^{+0.73}_{-0.15}$ & $0.29^{+1.19}_{-0.18}$ & $1.78^{+0.08}_{-0.08}$ & $0.59^{+2.36}_{-0.31}$\\
\hline				
100 & $0.21$ & $3675^{+3019}_{-2536}$ & $1.02^{+0.76}_{-0.34}$ & $0.47^{+1.44}_{-0.29}$ & $0.59^{+2.29}_{-0.34}$ & $1.78^{+0.56}_{-0.40}$ & $1.07^{+4.68}_{-0.68}$\\
100 & $0.84$ & $902^{+633}_{-635}$ & $0.51^{+0.44}_{-0.16}$ & $0.29^{+0.88}_{-0.18}$ & $0.34^{+1.44}_{-0.19}$ & $1.78^{+0.31}_{-0.27}$ & $0.68^{+2.87}_{-0.38}$\\
100 & $1.84$ & $403^{+315}_{-300}$ & $0.38^{+0.43}_{-0.13}$ & $0.25^{+0.80}_{-0.15}$ & $0.31^{+1.27}_{-0.18}$ & $1.78^{+0.17}_{-0.16}$ & $0.60^{+2.56}_{-0.32}$\\
100 & $\pi$ & $227^{+216}_{-184}$ & $0.33^{+0.46}_{-0.12}$ & $0.25^{+0.77}_{-0.15}$ & $0.31^{+1.24}_{-0.19}$ & $1.78^{+0.13}_{-0.16}$ & $0.60^{+2.49}_{-0.33}$\\
100 & $2\pi$ & $65.6^{+156.2}_{-38.3}$ & $0.29^{+0.48}_{-0.13}$ & $0.25^{+0.77}_{-0.15}$ & $0.31^{+1.24}_{-0.18}$ & $1.78^{+0.13}_{-0.12}$ & $0.59^{+2.50}_{-0.31}$\\
100 & $4\pi$ & $41.3^{+18.4}_{-15.3}$ & $0.29^{+0.48}_{-0.14}$ & $0.25^{+0.77}_{-0.15}$ & $0.30^{+1.24}_{-0.19}$ & $1.78^{+0.13}_{-0.12}$ & $0.60^{+2.49}_{-0.32}$\\
\hline				
\end{tabular}
\end{center}
\caption{Typical uncertainties in the measurement of the GW source parameters as a function of the total number of pulsars in the array $M$ and their sky coverage $\Delta \Omega_\mathrm{PTA}$ (the portion of the sky over which the pulsars are uniformly distributed). For each PTA configuration we consider $2.5\times10^4$--to--$1.6\times10^6$ (depending on the number of pulsars in the array) GW sources with random parameters. The GW source location is drawn uniformly in the sky, and the other parameters are drawn uniformly over the full range of $\psi$, $\phi_0$ and $\cos\iota$, $f_0$ is fixed at $5\times10^{-8}$Hz. In every Monte Carlo realisation, the optimal SNR is equal to 10. The table reports the median of the statistical errors $\Delta \lambda$ -- where $\lambda$ is a generic source parameter -- and the 25$^{{\rm th}}$ and 75$^{{\rm th}}$ percentile of the distributions obtained from the Monte Carlo samplings. Note that the errors $\Delta R/R$, $\Delta \iota$, $\Delta \psi$, $\Delta f$, $\Delta \Phi_0$ all scale as SNR$^{-1}$, the error $\Delta\Omega$ scales as SNR$^{-2}$.}
\label{tab:summary}
\end{table*}

In this section we present and discuss the results of our analysis aimed at determining the uncertainties surrounding the estimates of the GW source parameters. We focus in particular on the sky localization of a MBHB, which is of particular interest for possible identifications of electromagnetic counterparts, including the host galaxy and/or galactic nucleus in which the MBHB resides. For the case of binaries in circular orbit and whose gravitational radiation does not produce a measurable frequency drift, the mass and distance are degenerate, and can not be individually measured: one can only measure the combination ${\cal M}^{5/3}/D_L$. This prevents measurements of MBHB masses, which would be of great interest. On the other hand, the orientation of the orbital angular momentum -- through measurements of the inclination angle $\iota$ and the polarisation angle $\psi$ -- can be determined (although only with modest accuracy, as we will show below), which may be useful in determining the geometry of the system, if a counterpart is detected.

The uncertainties on the source parameters depend on a number of factors, including the actual MBHB parameters, the SNR, the total number of pulsars and their location in the sky with respect to the GW source. It is therefore impossible to provide a single figure of merit that quantifies of how well PTAs will be able to do GW astronomy. One can however derive some general trends and scalings, in particular how the results depend on the number of pulsars and their distribution in the sky, which we call the {\em sky coverage of the array}; this is of particular importance to design observational campaigns, and to explore tradeoffs in the observation strategy. In the following subsections, by means of extensive Monte Carlo simulations, we study the parameter estimation accuracy as a function of the number of pulsars in the array, the total SNR of the signal, and on the array sky coverage. All our major findings are summarised in Table \ref{tab:summary}.
  
\subsection{General behavior}

Before considering the details of the results we discuss conceptually the process by which the source parameters can be measured. Our discussion is based on the assumption that the processing of the data is done through a coherent analysis.The frequency of the signal is trivially measured, as this is the key parameter that needs to be matched in order for a template to remain in phase with the signal throughout the observation period. Furthermore, the amplitude of the GW signal determines the actual SNR, and is measured in a straightforward way. The amplitude $R$, or equivalently $A_\mathrm{gw}$, see Eqs.~(\ref{e:Agw}) and~(\ref{erre}), provides a constraint on the chirp mass and distance combination ${\cal M}^{5/3}/D_L$. However, in the case of monochromatic signals, these two parameters of great astrophysical interest can not be measured independently. If the frequency derivative $\dot{f}$ were also observable -- this case is not considered in this paper, as it likely pertains only to a small fraction of detectable binaries, see Section~\ref{s:signal} and Fig.~\ref{fig1b} -- then one would be able to measure independently both the luminosity distance and chirp mass. In fact, from the measurement of $\dot{f} \propto {\cal M}^{5/3} f^{11/3}$, that can be evaluated from the phase evolution of timing residuals, one can measure the chirp mass, which in turn, from the observation of the amplitude, would yield an estimate of the luminosity distance\footnote{We note that a direct measurement of the chirp mass would be possible if one could detect both the Earth- and pulsar-terms, \emph{provided that the distance to the pulsar was known}. In this case one has the GW frequency at Earth, the GW frequency at the pulsar, and the Earth-pulsar light-travel-time, which in turns provides a direct measure of $\dot{f}$, and as a consequence of the chirp mass.}. The remaining parameters, those that determine the geometry of the binary -- the source location in the sky, and the orientation of the orbital plane -- and the initial phase $\phi_0$ can be determined only if the PTA array contains at least three (not co-aligned) pulsars. The source location in the sky is simply reconstructed through geometrical triangulation, because the PTA signal for each pulsar encodes the source coordinates in the sky in the relative amplitude of the sine and cosine term of the response or, equivalently, the overall phase and amplitude of the sinusoidal PTA output signal, see Eqs.~(\ref{e:r}),~(\ref{e:z1}),~(\ref{e:FA}) and~(\ref{researth}). For the reader familiar with GW observations with {\it LISA}, we highlight a fundamental difference between {\it LISA} and PTAs in the determination of the source position in the sky. With {\it LISA}, the error box decreases as the signal frequency increases (everything else being equal), because the source location in the sky is reconstructed (primarily) through the location-dependent Doppler effect produced by the motion of the instrument during the observation, which is proportional to the signal frequency. This is not the case for PTAs, where the error-box is independent of the GW frequency. It depends however on the number of pulsars in the array -- as the number of pulsars increases, one has to select with increasingly higher precision the actual value of the angular parameters, in order to ensure that the same GW signal fits correctly the timing residuals of all the pulsars -- and the location of the pulsars in the sky.

\begin{figure}
\centerline{\psfig{file=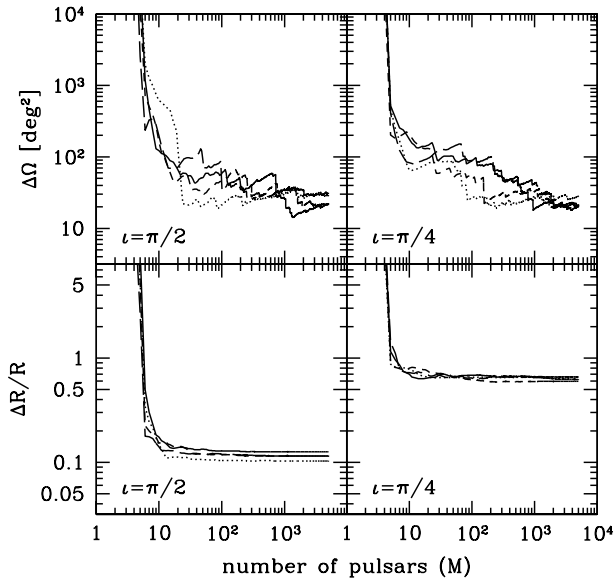,width=84.0mm}}
\caption{The statistical errors that affect the determination of the source location $\Delta\Omega$, see Eq.~(\ref{domega}) (upper panels) and the signal amplitude $R$ (lower panels) for four randomly selected sources (corresponding to the different line styles). We increase the number of pulsars in the array fixing a total SNR$=10$, and we plot the results as a function of the number of pulsars $M$. In the left panels we consider selected edge-on ($\iota=\pi/2$) sources, while in the right panel we plot sources with intermediate inclination of $\iota=\pi/4$.}
\label{fig2a}
\end{figure}

We first consider how the parameter estimation depends on the total number of pulsars $M$ at fixed SNR. We consider a GW source with random parameters and we evaluate the inverse of the Fisher information matrix as we progressively add pulsars to the array. The pulsars are added randomly from a uniform distribution in the sky and the noise has the same spectral density for each pulsar. We also keep the total coherent SNR fixed, at the value SNR = 10. It is clear that in a real observation the SNR actually increases approximately as $\sqrt{M}$, and therefore depends on the number of pulsars in the array. However, by normalising our results to a constant total SNR, we are able to disentangle the change in the uncertainty on parameter estimation that depends on the number of pulsars from the change due simply to the SNR. The results are shown in Fig. \ref{fig2a}. The main effect of adding pulsars in the PTA is to improve the power of triangulation and to reduce the correlation between the source parameters. At least three pulsars in the array are needed to formally resolve all the parameters; however, given the strong correlation in particular amongst $R$, $\iota$ and $\psi$ (which will be discussed later in more detail) a SNR $\sim100$ is needed to locate the source in the sky with an accuracy $\lesssim 50$ deg$^2$ in this case. It is clear that the need to maintain phase coherency between the timing residuals from several pulsars leads to a steep (by orders of magnitude) increase in accuracy from $M=3$ to $M\approx 20$ (note that the current Parkes PTA counts 20 pulsars). Adding more pulsars to the array reduces the uncertainty location region in the sky $\Delta \Omega$ by a factor of $\approx 5$ going from 20 to 1000 pulsars, but has almost no impact on the determination of the other parameters (the bottom panels of Fig. \ref{fig2a} show that $\Delta R/R$ is essentially constant for $M \simgt 20$).

\begin{figure}
\centerline{\psfig{file=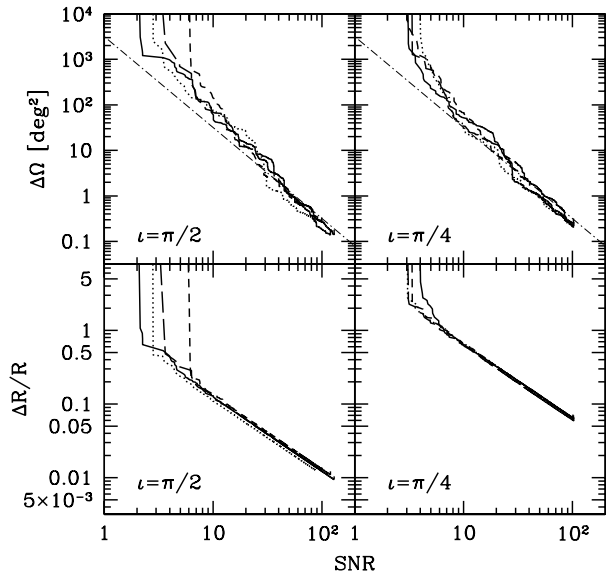,width=84.0mm}}
\caption{Same as Fig. \ref{fig2a}, but here, as we add pulsars to the PTA, we consistently take into account the effect on the total coherent SNR, and accordingly we plot the results as a function of the SNR. In the left panels we plot selected edge-on ($\iota=\pi/2$) sources, while in the right panel we consider selected sources with intermediate inclination of $\iota=\pi/4$. The dotted--dashed thin lines in the upper panels follow the scaling $\Delta\Omega \propto \mathrm{SNR}^{-2}$.}
\label{fig2b}
\end{figure}

Now that we have explored the effect of the number of pulsars alone (at fixed SNR) on the parameter errors, we can consider the case in which we also let the SNR change. We repeat the analysis described above, but now the SNR is not kept fixed and we let it vary self-consistently as pulsars are added to the array. The results plotted as a function of the total coherent SNR are shown in Fig. \ref{fig2b}. Once more, we concentrate in particular on the measurement of the amplitude $R$ and the error box in the sky $\Delta\Omega$. For $M \gg 1$, the error box in the sky and the amplitude measurements scale as expected according to  $\Delta \Omega\propto\mathrm{SNR}^{-2}$ and $\Delta R/R \propto \mathrm{SNR}^{-1}$ (and so do all the other parameters not shown here) . However, for $\mathrm{SNR} \simlt 10$ the uncertainties departs quite dramatically from the scaling above simply due to fact that with only a handful of pulsars in the array the strong correlations amongst the parameters degrade the measurements. We stress that the results shown here are independent of the GW frequency; we directly checked this property by performing several tests, in which the source's frequency is drawn randomly in the range $10^{-8}$ Hz - $10^{-7}$ Hz.

\begin{figure}
\centerline{\psfig{file=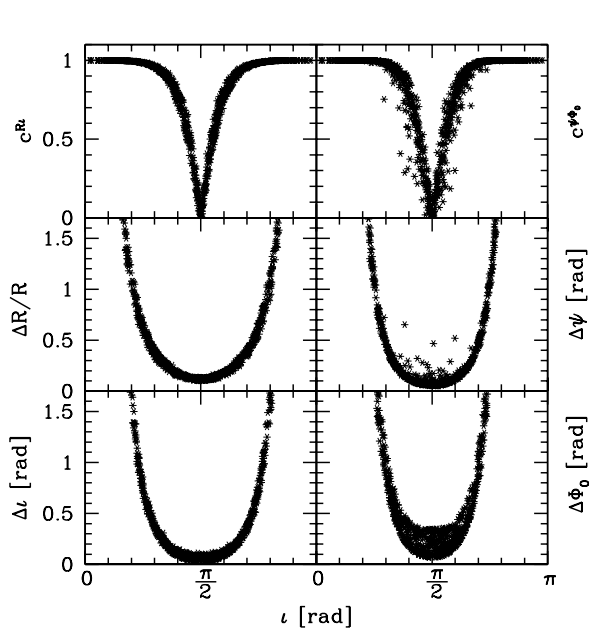,width=84.0mm}}
\caption{The effect of the source orbital inclination $\iota$ on the estimate of the signal parameters. Upper panels: The correlation coefficients $c^{R\iota}$ (left) and $c^{\psi\Phi_0}$ (right) as a function of $\iota$. Middle and bottom panels: the statistical errors in the measurement of amplitude $R$, polarisation angle $\psi$, inclination angle and initial phase $\Phi_0$ for a fixed PTA coherent SNR = 10, making clear the connection between inclination, correlation (degeneracy) and parameter estimation. Each asterisk on the plots is a randomly generated source.}
\label{fig3}
\end{figure}
\begin{figure}
\centerline{\psfig{file=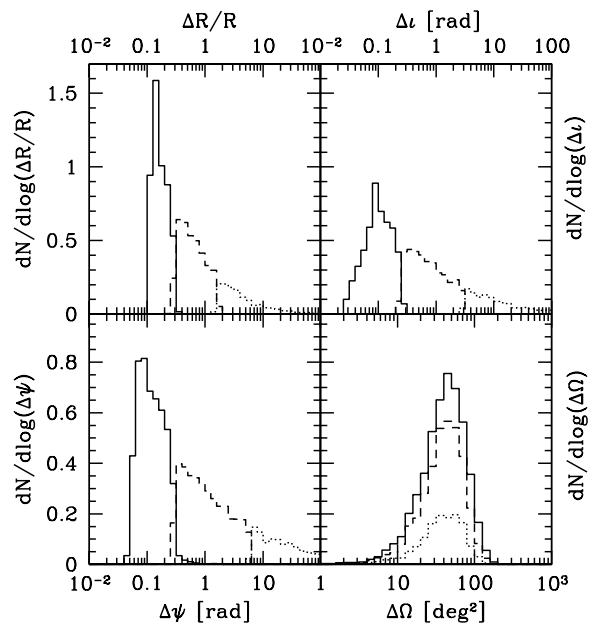,width=84.0mm}}
\caption{The distributions of the statistical errors of the source parameter measurements using a sample of 25000 randomly distributed sources (see text for more details), divided in three different inclination intervals: $\iota \in [0,\pi/6]\cup[5/6\pi,\pi]$ (dotted), $\iota \in [\pi/6,\pi/3]\cup[2/3\pi, 5/6\pi]$ (dashed) and $\iota\in [\pi/3, 2/3\pi]$ (solid). In each panel, the sum of the distribution's integrals performed over the three $\iota$ bins is unity.}
\label{fig4}
\end{figure}

The source inclination $\iota$ angle is strongly correlated with the signal amplitude $R$, and the polarisation angle $\psi$ is correlated to both $\iota$ and $\Phi_0$. The results are indeed affected by the actual value of the source inclination. Left panels in Figs. \ref{fig2a} and \ref{fig2b} refer to four different edge-on sources (i.e. $\iota=\pi/2$ and the radiation is linearly polarised). In this case, the parameter have the least correlation, and $\Delta R/R=$SNR$^{-1}$. Right panels in Figs. \ref{fig2a} and \ref{fig2b} refer to sources with an "intermediate" inclination  $\iota=\pi/4$; here degeneracies start to play a significant role and cause a factor of $\approx 3$ degradation in $\Delta R/R$ estimation (still scaling as SNR$^{-1}$). Note, however, that the sky position accuracy is independent on $\iota$ (upper panels in Figs. \ref{fig2a} and \ref{fig2b}), because the sky coordinates $\theta$ and $\phi$ are only weakly correlated to the other source parameters. We further explore this point by considering the behaviour of the correlation coefficients ($c^{R\iota}$ and $c^{\psi\Phi_0}$) as a function of $\iota$. Fig. \ref{fig3} shows the correlation coefficients and statistical errors in the source's parameters for a sample of 1000 individual sources using a PTA with $M=100$ and total SNR$=10$ as a function of $\iota$. For a face-on source ($\iota=0, \pi$), both polarizations equally contribute to the signal, and any polarization angle $\psi$ can be perfectly 'reproduced' by tuning the source phase $\Phi_0$, i.e. the two parameters are completely degenerate and cannot be determined. Moving towards edge-on sources, progressively change the relative contribution of the two polarizations, breaking the degeneracy with the phase. Fig. \ref{fig4}, shows statistical error distributions for the different parameters over a sample of 25000 sources divided in three different $\iota$ bins. The degradation in the determination of $R$, $\iota$ and $\psi$ moving towards face-on sources is clear. Conversely, both $\theta$ and $\phi$ do not have any strongly dependent correlation with the other parameters, the estimation of $\Omega$ is then independent on the source inclination (lower right panel in Fig. \ref{fig4}). 

\subsection{Isotropic distribution of pulsars}

\begin{figure}
\centerline{\psfig{file=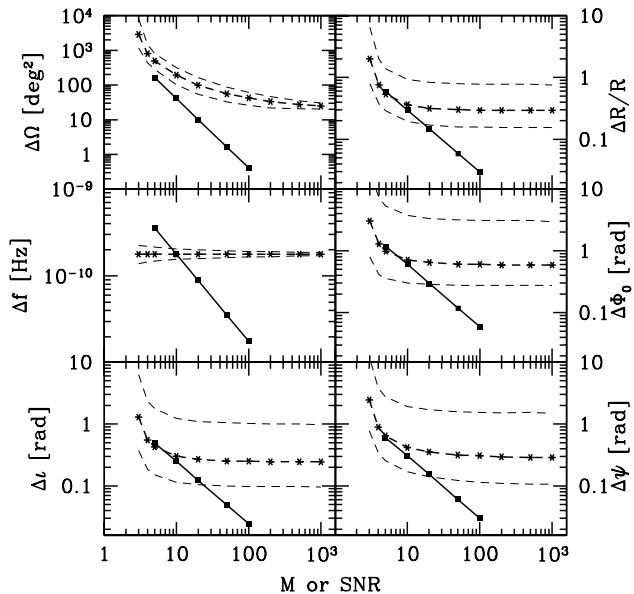,width=84.0mm}}
\caption{Median expected statistical error on the source parameters. Each point (asterisk or square) is obtained by averaging over a large Monte Carlo sample of MBHBs (it ranges from $2.5\times 10^4$ when considering 1000 pulsars to $1.6\times10^6$ when using 3 pulsars). In each panel, solid lines (squares) represent the median statistical error as a function of the total coherent SNR, assuming 100 randomly distributed pulsars in the sky; the thick dashed lines (asterisks) represent the median statistical error as a function of the number of pulsars $M$ for a fixed total SNR$=10$. In this latter case, thin dashed lines label the 25$^{\rm th}$ and the 75$^{\rm th}$ percentile of the error distributions.}
\label{fig5}
\end{figure}
\begin{figure}
\centerline{\psfig{file=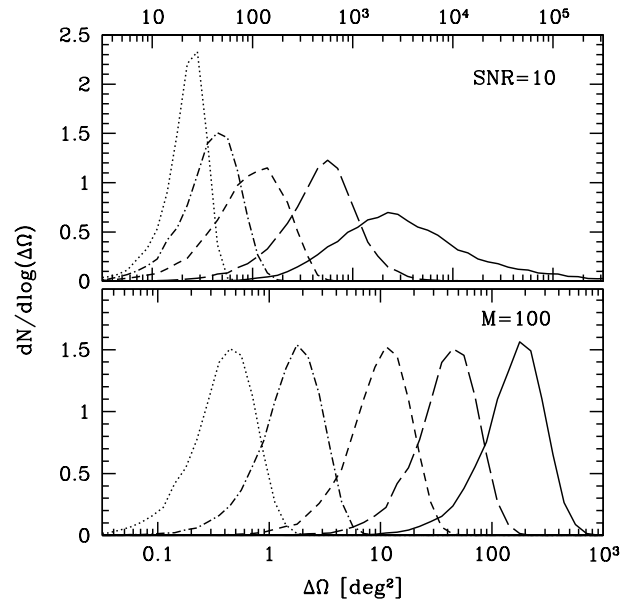,width=84.0mm}}
\caption{Distributions normalised to unity of the size of the error-box in the sky assuming an isotropic random distribution of pulsars in the array. Upper panel: from right to left the number of pulsars considered is $M=3, 5, 20, 100, 1000$, and we fixed a total SNR$=10$ in all cases. Lower panel: from right to left we consider SNR$=5, 10, 20, 50, 100$, and we fixed $M=100$.}
\label{fig6}
\end{figure}

In this Section we study the parameter estimation for a PTA whose pulsars are \emph{isotropically} distributed in the sky, and investigate how the results depend on the number $M$ of pulsars in the array and the SNR. Current PTAs have pulsars that are far from being isotropically located on the celestial sphere -- the anisotropic distribution of pulsars is discussed in the next Section -- but the isotropic case is useful to develop an understanding of the key factors that impact on the PTA performances for astronomy. It can also be considered representative of future PTAs, such as SKA, where many stable pulsars are expected to be discovered all over the sky.

We begin by fixing the total coherent SNR at which the GW signal is observed , and we set SNR$= 10$, regardless of the number of pulsars in the array, and explore the dependence of the results on the number of pulsars $M$ in the range 3-to-1000. We then consider a fiducial 'SKA-configuration' by fixing the total number of pulsars to $M=100$, and we explore how the results depend on the SNR for values $5 \le \mathrm{SNR} \le 100$. Throughout this analysis we assume that the timing noise is exactly the same for each pulsar and that the observations of each neutron star cover the same time span. The relative contribution of each of the pulsars in the PTA to the SNR is therefore solely dictated by the geometry of the system pulsar-Earth-source, that is the specific value of the beam patter function $F^{+,\times}(\theta, \phi,\psi)$. In total we consider 14 $M$-SNR combinations, and for each of them we generate $2.5\times 10^4$-to-$1.6\times10^6$ (depending on the total number of pulsars in the array) random sources in the sky. Each source is determined by the seven parameters described by Eq. (\ref{par}), which, in all the Monte Carlo simulations presented from now on, are chosen as follow. The angles $\theta$ and $\phi$ are randomly sampled from a uniform distribution in the sky; $\Phi_0$ and $\psi$ are drawn from a uniform distribution over their relevant intervals, [0,2$\pi$] and [0,$\pi$] respectively;$\iota$ is sampled according to a probability distribution $p(\iota)= \sin\iota/2$ in the interval $[0, \pi]$ and the frequency is fixed at $f=5\times 10^{-8}$ Hz. Finally the amplitude $R$ is set in such a way to normalise the signal to the pre-selected value of the SNR. For each source we generate $M$ pulsars randomly located in the sky and we calculate the Fisher information matrix and its inverse as detailed in Section~\ref{s:fim}. We also performed trial runs considering $f=10^{-7}$ Hz and $f=10^{-8}$ Hz (not shown here) to further cross-check that the results do not depend on the actual GW frequency.

Fig. \ref{fig5} shows the median statistical errors as a function of $M$ and SNR for all the six relevant source's parameters ($\theta$ and $\phi$ are combined into the single quantity $\Delta\Omega$, according to Eq.~(\ref{domega})). Let us focus on the $M$ dependence at a fixed SNR$=10$. The crucial astrophysical quantity is the sky location accuracy, which ranges from $\approx 3000$ deg$^2$ for $M=3$ -- approximately 10\% of the whole sky -- to $\approx 20$ deg$^2$ for $M=1000$. A PTA of 100 pulsars would be able to locate a MBHB within a typical error box of $\approx 40$ deg$^2$. The statistical errors for the other parameters are very weakly dependent on $M$ for $M\simgt 20$. The fractional error in the source amplitude is typically $\approx 30\%$, which unfortunately prevents to constrain an astrophysically meaningful ``slice'' in the ${\cal M}-D_L$ plane. The frequency of the source, which in this case was chosen to be $f = 5\times 10^{-8}$ Hz, is determined at a $\sim 0.1$ nHz level. Errors in the inclination and polarization angles are typically $\approx 0.3$ rad, which may provide useful information about the orientation of the binary orbital plane.

All the results have the expected scaling with respect to the SNR,  i.e. $\Delta\Omega \propto 1/\mathrm{SNR}^2$, and for all the other parameters shown in Fig.~\ref{fig5} the uncertainties scale as $1/\mathrm{SNR}$. A typical source with a SNR$=100$ (which our current astrophysical understanding suggests to be fairly unlikely, see SVV) would be located in the sky within an error box $\simlt 1\,\mathrm{deg}^2$ for $M \simgt 10$, which would likely enable the identification of any potential electro-magnetic counterpart. 

Distributions (normalised to unity) of $\Delta \Omega$ are shown in Fig. \ref{fig6}. The lower panel shows dependence on SNR (at fixed number of pulsars in the PTA, here set to 100), whose effect is to shift the distributions to smaller values of $\Delta \Omega$ as the SNR increases, without modifying the shape of the distribution. The upper panel shows the effectiveness of triangulation; by increasing the number of pulsars at fixed coherent SNR, not only the peak of the distribution shifts towards smaller values of $\Delta \Omega$, but the whole distribution becomes progressively narrower. If they yield the same SNR, PTAs containing a larger number of pulsars (sufficiently evenly distributed in the sky) with higher intrinsic noise are more powerful than PTAs containing fewer pulsars with very good timing stability, as they allow a more accurate parameter reconstruction (in particular for sky position) and they minimise the chance of GW sources to be located in "blind spots" in the sky (see next Section).

\subsection{Anisotropic distribution of pulsars}

\begin{figure}
\centerline{\psfig{file=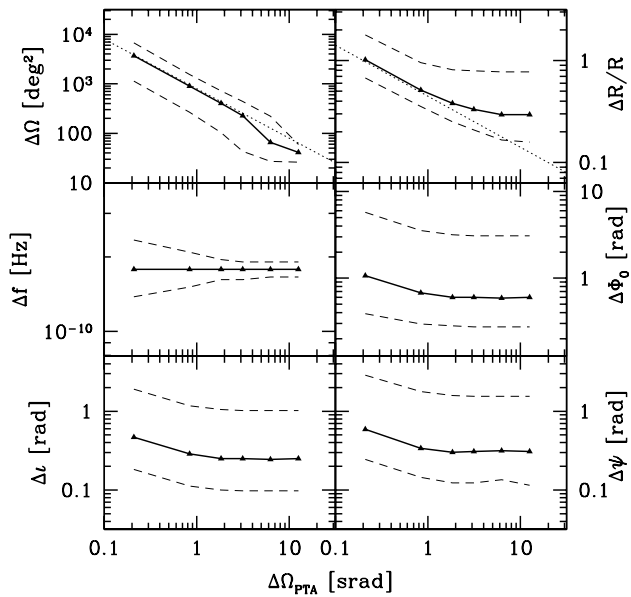,width=84.0mm}}
\caption{Median statistical error in the source's parameter estimation as a function of the sky-coverage of the pulsar distribution composing the array. Each triangle is obtained averaging over a Monte Carlo generated sample of $1.6\times10^5$ sources. In each panel, solid lines (triangles) represent the median error, assuming $M=100$ and a total SNR$=10$ in the array; thin dashed lines label  the 25$^{\rm th}$ and the 75$^{\rm th}$ percentile in the statistical error distributions.}
\label{fig7}
\end{figure}

\begin{figure*}
\centerline{\psfig{file=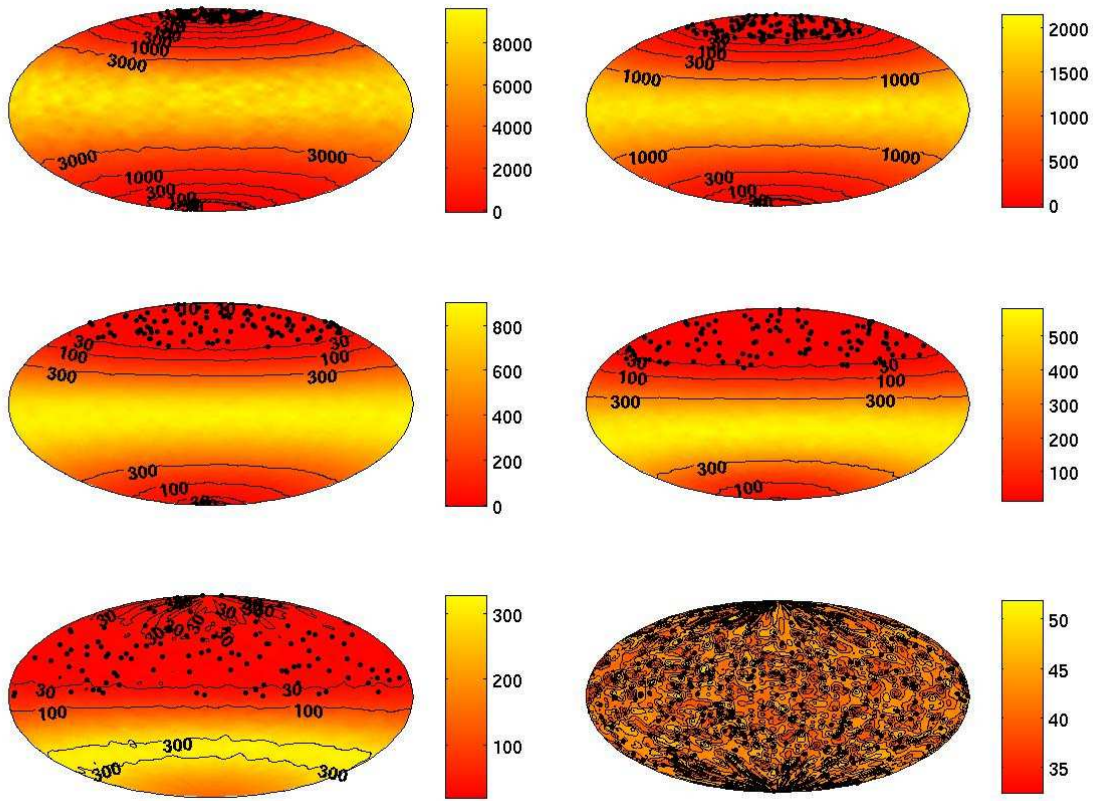,width=160.0mm}}
\caption{Sky maps of the median sky location accuracy for an anisotropic distribution of pulsars in the array. Contour plots are generated by dividing the sky into 1600 ($40\times40$) cells and considering all the randomly sampled sources falling within each cell; SNR$=10$ is considered. The pulsar distribution progressively fills the sky starting from the top left, eventually reaching an isotropic distribution in the bottom right panel (in this case, no distinctive features are present in the sky map). In each panel, 100 black dots label an indicative distribution of 100 pulsars used to generate the maps, to highlight the sky coverage. Labels on the contours refer to the median sky location accuracy expressed in square degrees, and the color--scale is given by the bars located on the right of each map.}
\label{fig8}
\end{figure*}

\begin{figure}
\centerline{\psfig{file=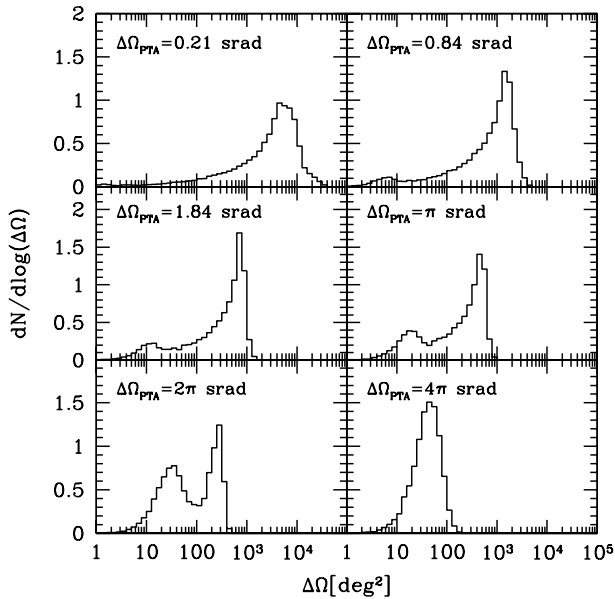,width=84.0mm}}
\caption{Normalized distributions of the statistical errors in sky position accuracy corresponding to the six sky maps shown in Fig. \ref{fig8}. Each distribution is generated using a random subsample of $2.5\times10^4$ sources.}
\label{fig9}
\end{figure}

\begin{figure*}
\centerline{\psfig{file=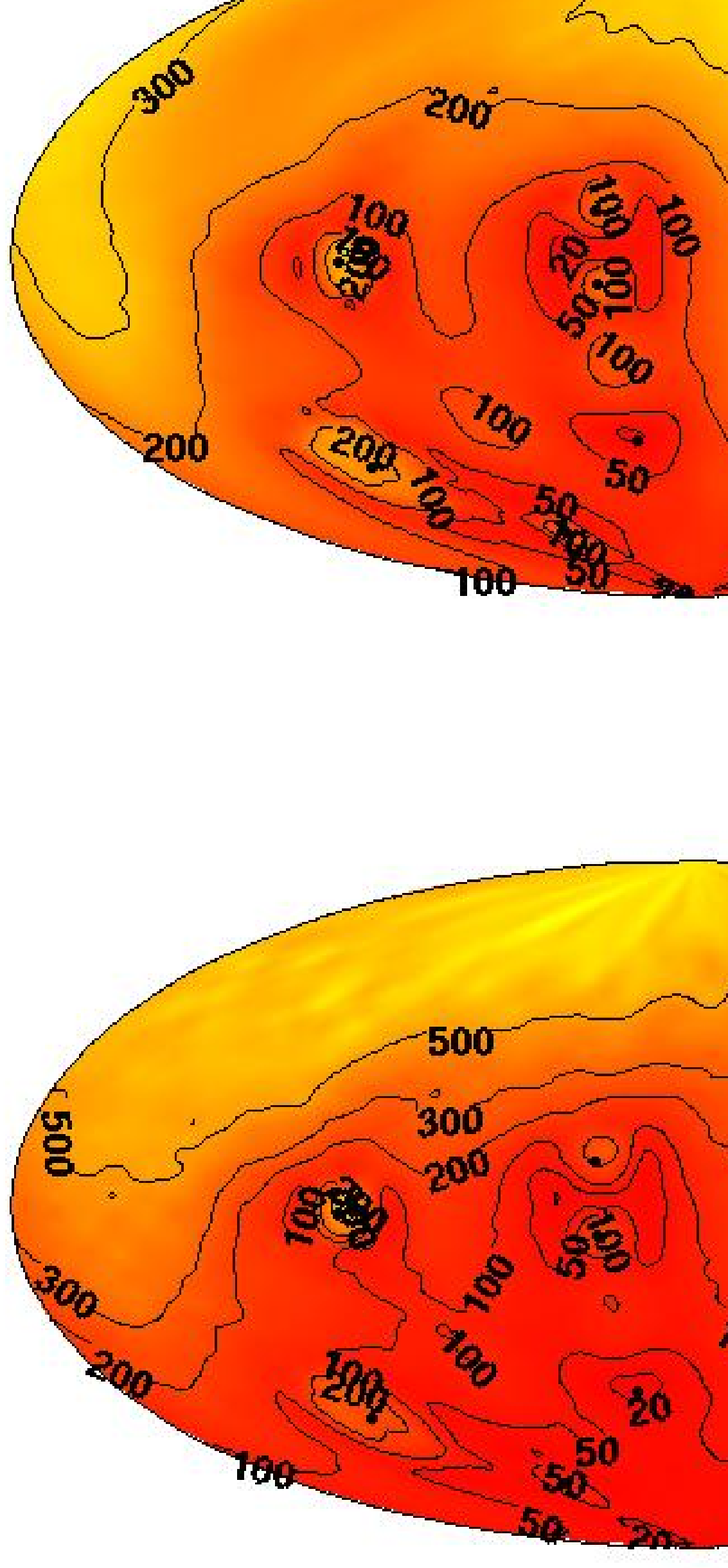,width=160.0mm}}
\caption{Sky maps of the median sky location accuracy for the Parkes PTA. Contour plots are generated as in Fig. \ref{fig8}. Top panel: we fix the source SNR$=10$ over the whole sky; in this case the sky position accuracy depends only on the different triangulation effectiveness as a function of the source sky location. Bottom panel: we fix the source chirp mass and distance to give a sky and polarization averaged SNR$=10$, and we consistently compute the mean SNR as a function of the sky position. The sky map is the result of the combination of triangulation efficiency and SNR as a function of the source sky location. The color--scale is given by the bars on the right, with solid angles expressed in deg$^2$.}
\label{fig10}
\end{figure*}

The sky distribution of the pulsars in a PTA is not necessarily isotropic. This is in fact the case for present PTAs, and it is likely to remain the norm rather than the exception, until SKA comes on-line. It is therefore useful -- as it also sheds new light on the ability of reconstructing the source parameters based on the crucial location of the pulsars of the array with respect to a GW source -- to explore the dependency of the results on what we call the "PTA sky coverage" $\Delta \Omega_\mathrm{PTA}$, i.e. the minimum solid angle in the sky enclosing the whole population of the pulsars in the array. We consider as a study case a `polar' distribution of 100 pulsars; the location in the sky of each pulsar is drawn from a uniform distribution in $\phi$ and $\cos\theta$ with parameters in the range $\phi \in [0,2\pi]$ and $\theta \in [0,\theta_{{\rm max}}]$, respectively. We then generate a random population of GW sources in the sky and proceed exactly as we have described in the previous section. We consider six different values of  $\Delta \Omega_\mathrm{PTA}$, progressively increasing the sky coverage. We choose $\theta_\mathrm{max} = \pi/12, \pi/6, \pi/4, \pi/3, \pi/2, \pi$ corresponding to $\Delta \Omega_\mathrm{PTA}=0.21, 0.84, 1.84, \pi, 2\pi, 4\pi$ srad. As we are interested in investigating the geometry effects, we fix in each case the total optimal SNR to 10. We dedicate the next section to consider specifically the case of the 20 pulsars that are currently part of the Parkes PTA.

The median statistic errors on the source parameters as a function of the PTA sky coverage are shown in Fig. \ref{fig7}. As one would expect, the errors decrease as the sky coverage increases, even if the SNR is kept constant. This is due to the fact that as the pulsars in the array populate more evenly the sky, they place increasingly more stringent constraints on the relative phase differences amongst the same GW signal measured at each pulsar, which depends on the geometrical factors $F^{+,\times}$. The most important effect is that the sky position is pinned down with greater accuracy; at the same time, correlations between the sky location parameters and other parameters, in particular amplitude and inclination angle are reduced. $\Delta \Omega$ scales linearly (at fixed SNR) with $\Delta \Omega_\mathrm{PTA}$, but the others parameters do not experience such a drastic improvement. The statistical uncertainty on the amplitude improves as $\sqrt{\Delta \Omega_\mathrm{PTA}}$ for $\Delta \Omega_\mathrm{PTA} \simlt 1$ srad, then saturates. All the other parameters are much less sensitive to the sky coverage, showing only a mild improvement (a factor $\lesssim 2$) with increasing $\Delta \Omega_\mathrm{PTA}$ up to $\sim 1$ srad. 

When one considers an anisotropic distribution of pulsars, the median values computed over a random uniform distribution of GW sources in the sky do not carry however the full set of information. In particular the error-box in the sky strongly depends on the actual source location. To show and quantify this effect, we use the outputs of the Monte Carlo runs to build sky maps of the median of $\Delta \Omega$ that we shown in Fig. \ref{fig8}. When the pulsars are clustered in a small $\Delta \Omega_\mathrm{PTA}$, the properties of the signals coming from that spot in the sky (and from the diametrically opposite one) are more susceptible to small variations with the propagation direction (due to the structure of the response functions $F^{+}$ and $F^{\times}$); the sky location can then be determined with a much better accuracy, $\Delta \Omega \sim 2$ deg$^2$. Conversely, triangulation is much less effective for sources located at right angles with respect to the bulk of the pulsars. For a polar $\Delta \Omega_\mathrm{PTA}=0.21$ srad, we find a typical $\Delta \Omega \gtrsim 5000$ srad for equatorial sources; i.e., their sky location is basically undetermined. Increasing the sky coverage of the array, obviously mitigates this effect, and in the limit $\Delta \Omega_\mathrm{PTA}=4\pi$ srad (which correspond to an isotropic pulsar distribution), we find a smooth homogeneous skymap without any recognizable feature (bottom right panel of Fig. \ref{fig8}). In this case the sky location accuracy is independent on the source sky position and, for $M = 100$ and $\mathrm{SNR} = 10$ we find $\Delta \Omega \sim 40$ deg$^2$. Fig. \ref{fig9} shows the normalized distributions of the statistical errors corresponding to the six skymaps shown in Fig. \ref{fig8}. It is interesting to notice the bimodality of the distribution for intermediate values of $\Delta \Omega_\mathrm{PTA}$, due to the fact that there is a sharp transition between sensitive and non sensitive areas in the sky (this is particularly evident looking at the contours in the bottom left panels of Fig. \ref{fig8}). 

We also checked another anisotropic situation of potential interest: a distribution of pulsars clustered in the Galactic plane. We considered a distribution of pulsars covering a ring in the sky, with $\phi_\alpha$ is randomly sampled in the interval [0,$2\pi$] and latitude in the range [$-\pi/12, \pi/12$] around the equatorial plane, corresponding to a solid angle of $\Delta \Omega_\mathrm{PTA}=3.26$ srad. Assuming a source SNR$=10$, the median statistical error in the source sky location is $\sim 100$ deg$^2$, ranging from $\sim 10$ deg$^2$ in the equatorial plane, to $\sim 400$ deg$^2$ at the poles. Median errors on the other parameters are basically the same as in the isotropic case.      

\subsection{The Parkes Pulsar Timing Array}

We finally consider the case that is most relevant to present observations: the potential capabilities of the Parkes Pulsar Timing Array. The goal of the survey is to monitor 20 milli-second pulsars for five years with timing residuals $\approx 100$ ns~\cite{man08}. This may be sufficient to enable the detection of the stochastic background generated by the whole population of MBHBs~\cite{papI}, but according to our current astrophysical understanding (see SVV) it is unlikely to lead to the detection of radiation from individual resolvable MBHBs, although there is still a non-negligible chance of detection. It is therefore interesting to investigate the potential of such a survey.

In our analysis we fix the location of the pulsars in the PTA to the coordinates of the 20 milli-second pulsars in the Parkes PTA, obtained from~\cite{ATNF-catalogue}; however for this exploratory analysis we set the noise spectral density of the timing residuals to be the same for each pulsar, \emph{i.e.} we do not take into account the different timing stability of the pulsars. We then generate a Monte Carlo sample of GW sources in the sky with the usual procedure. We consider two different approaches. Firstly, we explore the parameter estimation accuracy as a function of the GW source sky location for selected fixed array coherent SNRs (5, 10, 20, 50 and 100). Secondly, we fix the source chirp mass, frequency and distance (so that the sky and polarization averaged coherent SNR is 10) and we explore the parameter estimation accuracy as a function of the sky location. Skymaps of statistical error in the sky location are shown in Fig. \ref{fig10}. In the top panel we fix the SNR$=10$, independently on the source position in the sky; the median error in the sky location accuracy is $\Delta \Omega \sim 130$ deg$^2$, but it ranges from $\sim 10$ deg$^2$ to $\sim400$ deg$^2$ depending on the source's sky location. The median statistical errors that affect the determination of all the other source parameters are very similar to those for the isotropic pulsar distribution case when considering $M=20$, since the pulsar array covers almost half of the sky, see Fig. \ref{fig7}. In the bottom panel, we show the results when we fix the source parameters, and therefore, the total SNR in the array does depend on the source sky location. In the southern hemisphere, where almost all the pulsars are concentrated, the SNR can be as high as 15, while in the northern hemisphere it can easily go below 6. The general shape of the skymap is mildly affected, and shows an even larger imbalance between the two hemispheres. In this case, the median error is $\Delta \Omega \sim 160$ deg$^2$, ranging from $\sim 3$ deg$^2$ to $\sim900$ deg$^2$. It is fairly clear that adding a small ($\simlt 10$) number pulsars in the northern hemisphere to the pulsars already part of the Parkes PTA would significantly improve the uniformity of the array sensitivity and parameter estimation capability, reducing the risk of potentially detectable GW sources ending up in a "blind spot" of the array.

\section{Conclusions}

In this paper we have studied the expected uncertainties in the measurements of the parameters of a massive black hole binary systems by means of gravitational wave observations with Pulsar Timing Arrays. We have investigated how the results vary as a function of the signal-to-noise ratio, the number of pulsars in the array and their location in the sky with respect to a gravitational wave source. Our analysis is focused on MBHBs in circular orbit with negligible frequency evolution during the observation time ("monochromatic sources"), which we have shown to represent the majority of the observable sample, for sensible models of sub--parsec MBHB eccentricity evolution. The statistical errors are evaluated by computing the variance-covariance matrix of the observable parameters, assuming a coherent analysis of the Earth-terms only produced by the timing residuals of the pulsars in the array (see Section II B).

For a fiducial case of an array of 100 pulsars randomly distributed in the sky, assuming a coherent total SNR = 10, we find a typical error box in the sky $\Delta \Omega \approx 40$ deg$^2$ and a fractional amplitude error of $\approx 0.3$. The latter places only very weak constraints on the chirp mass-distance combination ${\cal M}^{5/3}/D_L$. At fixed SNR, the typical parameter accuracy is a very steep function of the number of pulsars in the PTA up to $\approx 20$. For PTAs containing more pulsars, the actual gain becomes progressively smaller because the pulsars ``fill the sky" and the effectiveness of further triangulation weakens. We also explored the impact of having an anisotropic distribution of pulsars finding that the typical source sky location accuracy improves linearly with the array sky coverage. For the specific case of the Parkes PTA where all the pulsars are located in the southern sky, the sensitivity and sky localisation are significantly better (by an order of magnitude) in the southern hemisphere, where the error-box is $\lesssim 10 \,\mathrm{deg}^2$ for a total coherent SNR = 10. In the northern hemisphere, the lack of monitored pulsars prevent a source location to be in an uncertainty region $\lesssim 200\,\mathrm{deg}^2$. The monitoring of a handful of pulsars in the northern hemisphere would significantly increase both the SNR and the parameter recovery of GW sources, and the International PTA~\cite{HobbsEtAl:2009} will provide such a capability in the short term future.

The main focus of our analysis is on the sky localisation because sufficiently small error-boxes in the sky may allow the identification of an electro-magnetic counterpart to a GW source. Even for error-boxes of the order of tens-to-hundreds of square degrees (much larger than  \emph{e.g.} the typical {\it LISA} error-boxes~\cite{v04,k07,lh08}), the typical sources are expected to be massive (${\cal M} \simgt 10^{8}\msun$) and at low redshift ($z\simlt 1.5$), and therefore the number of associated massive galaxies in the error-box should be limited to a few hundreds. Signs of a recent merger, like the presence of tidal tails or irregularities in the galaxy luminosity profile, may help in the identification of potential counterparts. Furthermore, if nuclear activity is present, \emph{e.g.} in form of some accretion mechanism, the number of candidate counterparts would shrink to an handful, and periodic variability \cite{hkm07} could help in associating the correct galaxy host. We are currently investigating the astrophysical scenarios and possible observational signatures, and we plan to come back to this important point in the future. The advantage of a counterpart is obvious: the redshift measurement would allow us, by assuming the standard concordance cosmology, to measure the luminosity distance to the GW source, which in turn would break the degeneracy in the amplitude of the timing residuals $R \propto {\cal M}^{5/3}/(D_L f^{1/3})$ between the chirp mass and the distance, providing therefore a direct measure of ${\cal M}$.

The study presented in this paper deals with monochromatic signals. However, the detection of MBHBs which exhibit a measurable frequency drift would give significant payoffs, as it would allow to break the degeneracy between distance and chirp mass, and enable the direct measurement of both parameters. Such systems may be observable with the Square-Kilometre-Array. In the future, it is therefore important to extend the present analysis to these more general signals. However, as the frequency derivative has  only modest correlations with the sky position parameters, we expect that the results for the determination of the error-box in the sky discussed in this paper will still hold. A further extension to the work is to consider MBHBs characterised by non-negligible eccentricity, which is currently in progress. Another extension to our present study is to consider both the Earth- and pulsar-terms in the analysis of the data and the investigation of the possible benefits of such scheme, assuming that the pulsar distance is not known to sufficient accuracy. This also raises the issue of possible observation campaigns that could yield an accurate (to better than 1 pc) determination of the pulsar distances used in PTAs. In this case the use of the pulsar-term in the analysis would not require the introduction of (many more) unknown parameters and would have the great benefit of breaking the degeneracy between chirp mass and distance.

The final world of caution goes to the interpretation of the results that we have presented in the paper. The approach based on the computation of the Fisher Information matrix is powerful and straightforward, and is justified at this stage to understand the broad capabilities of PTAs and to explore the impact on astronomy of different observational strategies. However, the statistical errors that we compute are strictly \emph{lower limits} to the actual errors obtained in a real analysis; the fact that at least until SKA comes on line, a detection of a MBHB will be at a moderate-to-low SNR should induce caution in the way in which the results presented here are interpreted. Moreover, in our current investigation, we have not dealt with a number of important effects that in real life play a significant role, such as different calibrations of different data sets, the change of systematic factors that affect the noise, possible non-Gaussianity and non-stationarity of the noise, etc. These (and other) important issues for the study of MBHBs with PTAs should be addressed more thoroughly in the future by performing actual mock analyses and developing suitable analysis algorithms.

\appendix
\section{The pulsar timing array response to gravitational waves}

In this Appendix we derive the PTA response to a GW signal from a deterministic source characterised by a metric perturbation 
\be
h(t,\vec{x}_0) = \left[A_+(t,\vec{x}_0) F^+(\ho)  - i A_\times(t,\vec{x}_0) F^\times(\ho)\right]  e^{i\phi(t)}\,.
\ee
The observed signal at the PTA output generated by the fractional frequency shift comes from the contribution at the "Earth" (or more precisely the solar system barycentre, SSB) and at the pulsar. We therefore need to compute the metric perturbation at $(t_\mathrm{ssb}, \vec{x}_\mathrm{ssb})$ and  $(t_\mathrm{p}, \vec{x}_\mathrm{p})$. For simplicity (but without loss of generality), let us make a choice of coordinates such that:
\ba
&& t_\mathrm{ssb} = t \,\quad \vec{x}_\mathrm{ssb} = 0\,,
\nonumber\\
&& t_\mathrm{p} = t-\tau \,\quad \vec{x}_\mathrm{p} = \vec{x}_\mathrm{p}\,,
\ea
where
\be
\tau = L \left(1 + \hat\Omega \cdot \hat{p}\right)\,.
\label{e:tau}
\ee
is the time delay between the pulsar and the Earth terms, given by the light-travel-time. The relative frequency shift at time $t$, Eq.~(\ref{e:z}) is therefore
\ba
z(t,\ho) & = & \left[A_+(t - \tau) F^+(\ho) - i A_\times(t - \tau) F^\times(\ho)\right] e^{i\phi(t - \tau)}
\nonumber\\
&& - \left[A_+(t) F^+(\ho) - i A_\times(t) F^\times(\ho)\right] e^{i\phi(t)}
\ea
As one expects, the response to a GW is identically zero for GWs propagating along the Earth-pulsar direction, when $\ho = \pm \hat{p}$. In fact, when GWs propagate in the direction opposite to the radio signal, $\ho \cdot \hat{p} = 1$, we have $\hm \cdot \hp  =  \hn \cdot \hp = 0$ and as a consequence $F^+(\ho) = F^\times(\ho) = 0$. This is due to the transverse nature of GWs and there is no frequency shift of the electro-magnetic wave. On the other hand, when GWs propagate along the same direction of the electromagnetic waves, $\ho \cdot \vec{p} = - 1$ and $F^+(\ho) = \cos(2\psi)$ and  $F^\times(\ho) = \sin(2\psi)$ are non-zero, finite functions of the polarization angle $\psi$. In this case $z(t,\ho)$ is still zero because $\tau = 0$ and the Earth and pulsar terms are identical and cancel out. This is known as the \emph{surfing effect}, as the GWs surf with the electro-magnetic waves.

We have shown in Section~\ref{s:signal} that GWs emitted by MBHBs and observable with PTA will be quasi-monochromatic signals slowly drifting in time. More specifically, for the astrophysical population that one expects to detect (see SVV), the frequency derivative satisfies (for the vast majority of the signals) the condition
\be
T \ll \frac{f_0}{\dot{f}_0} \ll L\,.
\ee
The consequence of the inequality above is that the timing residuals observed at the PTA output from any pulsar $\alpha$ will consist of two quasi-monochromatic signals at different, but essentially constant frequencies during the observation time $f(t)$ and $f(t - \tau_\alpha)$, where we have explicitly emphasised with $\tau_\alpha$ the fact that the pulsar-term differs from pulsar to pulsar, and will be sitting at a different frequency determined by the source-Earth-pulsar relative angle $\ho \cdot \hat{p}_\alpha$ and the \emph{ poorly constrained} pulsar distance $L_\alpha$. Given then the fact that the noise dominates the signal, $r_\alpha \ll n_\alpha$, only adding the contributions to all the pulsars will provide enough SNR, and one can therefore concentrate the analysis only on the Earth-terms, all of which depend only on the 7 parameters of the source, and ignore the contribution from the pulsar terms. The latter in fact depend on the $M$ unknown parameters $L_\alpha$'s, which determine $\tau_\alpha$ and therefore phase, frequency and amplitude of the pulsar term. If $L_\alpha$'s were known -- an uncertainty smaller than 1 rad on the phase contribution requires the distance to a pulsar to be known to better than $\approx 0.1\, (f/10\,\mathrm{nHz})^{-1}\,(1 + \hat\Omega \cdot \hat{p})^{-1}$ pc -- one can coherently lock all the phases of the pulsar terms increasing the total SNR of the detection. This would also provide a direct measurement of $\dot{f}$, allowing to break the distance--chirp mass degeneracy. 

The pulsar-term however may conjure to cancel the Earth-term for specific source-Earth-pulsar angles. To have an order of magnitude estimate of this effect, let us conservatively assume that the Earth and pulsar term can be fully resolved if their frequency separation is larger than one frequency resolution bin of width $1/T$ (we have actually shown in Section~\ref{s:results} that the frequency can be resolved with sub-frequency resolution accuracy). Assuming (again conservatively) a linear frequency shift due to radiation reaction, this condition can be expressed as
\be
\dot{f}_0 \tau > \frac{1}{T}\,.
\ee
Substituting Eq.~(\ref{e:tau}) in the previous expression, the previous condition can be expressed as
\be
1 + \ho \cdot \hat{p} > \frac{1}{T\dot{f}_0 L} > \frac{1}{f_0L}\,.
\ee
The cancellation takes place when $\ho \cdot \hat{p} \simeq 1$ and we can therefore approximate $1 + \ho \cdot \hat{p}$ as
\be
1 + \ho \cdot \hat{p} = 1 + \cos (\pi - \delta\theta) = \frac{1}{2} \delta\theta^2 + {\cal O}(\delta\theta^4)\,,\quad \delta\theta \ll 1\,.
\ee
Using a typical pulsar distance of 1 kpc and a GW frequency of $10$ nHz we obtain:
\be
\delta\theta \simgt 3 \left(\frac{10\,\mathrm{nHz}}{f_0}\right)^{1/2}\,\left(\frac{1\,\mathrm{kpc}}{L}\right)^{1/2}\,\mathrm{deg}\,.
\label{e:dteta}
\ee
This means that whenever the separation on the sky of a GW source and  a pulsar is larger than a few degrees, then ignoring the contribution from the pulsar term does not affect the analysis. With existing PTAs, this is surely a safe assumption unless one encounters a very unlucky case. Nonetheless, we keep this into account in the Monte Carlo simulations whose results are reported in Section~\ref{s:results}. We choose $50$nHz for the source frequency. Then, for every PTA array configuration, we place all the pulsars either at 1 kpc or at 5 kpc (in Section~\ref{s:results} we show results for $L_\alpha=5$ kpc, but the results for $L_\alpha=1$ kpc are basically identical); while for the Parkes PTA we took each individual pulsar distance from the ATNF catalogue \cite{ATNF-catalogue}. Finally, for each pulsar we compute the pulsar-Earth-source angle and if the resulting value is smaller than the one obtained from Eq.~(\ref{e:dteta}) we discard that particular pulsar contribution from the analysis; as $\delta\theta$ is very small, the impact on the analysis is minimal.

\end{document}